\documentclass[12pt]{article}
\pdfoutput=1
\usepackage[margin=3cm]{geometry}
\usepackage[utf8]{inputenc}
\usepackage{amsmath}
\usepackage{amssymb}
\usepackage{authblk} 
\usepackage{xcolor}
\usepackage{physics}
\usepackage{graphicx} 
\usepackage{slashed}
\usepackage{hyperref}
\usepackage{varioref}       
\usepackage{cleveref}    
\usepackage{soul}

\crefname{table}{table}{tables}
\Crefname{table}{Table}{Tables}
\crefname{figure}{figure}{figures}
\Crefname{figure}{Figure}{Figures}

\definecolor{darkred}{rgb}{0.7, 0.0, 0.2}
\definecolor{tealblue}{rgb}{0.21, 0.56, 0.63}
\hypersetup{colorlinks=true,allcolors = darkred,linktocpage=true}

\usepackage{pgf} 

\usepackage[
        backend=bibtex,
        sorting=none,
        style=phys,
        eprint=true,
        doi=false,
        biblabel=brackets
        ]{biblatex}

\bibliography{hs_cft.bib}

\numberwithin{equation}{section}
\numberwithin{table}{section}

\newenvironment{eqaed}
    {\begin{equation}
    \begin{aligned}
    }
    { 
    \end{aligned}
    \end{equation}
    \ignorespacesafterend
    }

\usepackage{setspace}

\begin{document}

\title{Infinite distances in multicritical CFTs\\ and higher-spin holography}

\author{Ivano Basile}
\author{Andrea Campoleoni}
\author{Simon Pekar}
\author{Evgeny Skvortsov\footnote{Also on leave from Lebedev Institute of Physics, Moscow, Russia.}}
\affil{\emph{Service de Physique de l'Univers, Champs et Gravitation}\\\emph{Universit\'{e} de Mons -- UMONS, Place du Parc 20, 7000 Mons, Belgium}}

\date{}

\maketitle

\begin{center}

\vspace{-45pt}
{\tt\small 
\{\href{mailto:ivano.basile@umons.ac.be}{ivano.basile},
\href{mailto:andrea.campoleoni@umons.ac.be}{andrea.campoleoni},
\href{mailto:simon.pekar@umons.ac.be}{simon.pekar},
\href{mailto:evgeny.skvortsov@umons.ac.be}{evgeny.skvortsov}\}@umons.ac.be
}

\end{center}

\vspace{30pt}

\begin{abstract}
    
     We investigate the swampland distance conjecture in higher-spin gravity. To this end, we study multicritical generalizations of large-$N$ vector models, bosonic and fermionic, and we compute the quantum information distance along selected renormalization-group trajectories toward the higher-spin limit. In striking contrast to the expected exponential decay of higher-spin masses or anomalous dimensions, we find that infinite-distance limits in these models lead to a power-like decay. This suggests that stringy exponential decays are characteristic of matrix-like gauge theories, rather than vector models. We corroborate this notion studying the information distance along coupling variations in Chern-Simons-matter CFTs, where matrix-like degrees of freedom dominate over vector-like ones. 
    
\end{abstract}
\newpage
\tableofcontents
\newpage

\section{Introduction}\label{sec:introduction}

In the absence of a 
clear-cut ultraviolet (UV) completion of gravity, it has proven fruitful to seek and identify general, model-independent principles that any theory of quantum gravity ought to satisfy. This is the ultimate aim of the swampland program~\cite{Vafa:2005ui} (see~\cite{Brennan:2017rbf, Palti:2019pca, vanBeest:2021lhn, Grana:2021zvf} for reviews), which has hitherto constructed a network of principles to distinguish theories of gravity that can be consistently UV-completed from those that cannot. These principles are largely conjectural to a varying degree, although some proofs in restricted contexts are available. Their mutual reinforcement and consistency suggests that they are pointing in the correct general direction, but so far, top-down evidence toward these proposals is almost completely found in string-theoretic constructions (see, however,~\cite{Basile:2021krr}). A natural avenue to explore in this context is higher-spin gravity~\cite{Bekaert:2022poo}, which may offer an alternative construction of consistent UV-complete theories. Whether higher-spin gravity is part of the string landscape remains an outstanding open problem. In high-dimensional supergravity, evidence is accumulating to the effect that string theory covers the entirety of theories satisfying swampland constraints
~\cite{Kim:2019ths, Katz:2020ewz, Montero:2020icj, Bedroya:2021fbu, Tarazi:2021duw}, but higher-spin gravity offers a different setting where these techniques may not be directly applicable.

In order to explore higher-spin gravity from the point of view of the swampland program, one can turn to holography, where higher-spin symmetry appears in a large-$N$ limit of (Chern-Simons) vector models. The swampland distance (or duality) conjecture~\cite{Ooguri:2006in} comes to mind: in its most general form, limiting regimes in a family of theories of gravity result in the emergence of a (possibly dual) weakly coupled description controlled by an infinite tower of light states. The namesake of the conjecture originates from the fact that the most well-understood examples of such families of theories are parameterized by moduli spaces, over which usually a canonical Riemannian metric is available. Limiting regimes are then identified as \emph{infinite-distance limits}. The distance conjecture has been refined in a number of directions (see, \emph{e.g.},~\cite{Baume:2016psm, Klaewer:2016kiy, Blumenhagen:2018nts, Heidenreich:2018kpg}), most notably the emergent string proposal~\cite{Lee:2018urn, Lee:2019wij, Lee:2019xtm}, stating that the leading light tower comprises either Kaluza-Klein states or higher-spin excitations of a tensionless critical string.

The latter is certainly relevant for any investigation of the higher-spin swampland, but the most appropriate version of the distance conjecture for the purposes of the present paper is its holographic incarnation, often dubbed the conformal field theory (CFT) distance conjecture~\cite{Baume:2020dqd, Perlmutter:2020buo}. This particular proposal lies entirely within the CFT side of the holographic duality, and is thus particularly convenient to discuss higher-spin gravity via large-$N$ vector models coupled to Chern-Simons theories. Concretely, the conjecture  
mainly consists in the equivalence between higher-spin symmetry and infinite-distance loci in the Zamolodchikov metric on a conformal manifold. This is indeed the dual notion of the metric on moduli spaces defined by non-linear sigma models of scalar fields in the gravitational bulk side. This formulation makes for a remarkably precise statement of the conjecture, but other cases of interest fall short of its purview due to the absence of a proper conformal manifold. For instance, non supersymmetric string constructions cannot readily be tackled with this approach (see, however,~\cite{Basile:2022zee}), and neither can the holographic duals of large-$N$ limits of vector models.

A natural extension of the Zamolodchikov metric that can be employed to circumvent this issue is the (quantum) \emph{information metric}~\cite{oconnor1993geometry, Dolan:1997cx, Stout:2021ubb, Stout:2021ubb}, where infinite-distance limits universally translate into factorization of correlators~\cite{Stout:2022phm}. In this paper we consider this generalized setting to define a distance for large-$N$ limits via renormalization group (RG) flows and studying the fate of higher-spin points with respect to the information metric. 
There are a number of holographic proposals regarding higher-spin gravity, including the breaking and recovery of higher-spin symmetry~\cite{Girardello:2002pp, Bianchi:2004xi, Gaberdiel:2015uca} which is relevant for our discussion. We will concentrate on the higher-spin/vector-model AdS/CFT duality~\cite{Klebanov:2002ja,Sezgin:2003pt,Leigh:2003gk,Giombi:2011kc,Bekaert:2012ux,Giombi:2012ms, Giombi:2016ejx}, although there is also an interesting relation between massless higher spin fields and tensionless strings~\cite{Sundborg:2000wp,Mikhailov:2002bp,Sezgin:2002rt} with a lot of progress on the string theory side over the last years~\cite{Gaberdiel:2014cha, Gaberdiel:2015mra,Gaberdiel:2015uca,Gaberdiel:2015wpo,Gaberdiel:2018rqv,Gaberdiel:2021qbb}. There is also a scenario~\cite{Chang:2012kt} with string theory on $\rm{AdS}_4\times \mathbb{CP}^3$ vs. ABJ models that allows one to interpolate between matrix- and vector-like duality. Finally, we would like to mention another approach based on matrix-model constructions~\cite{Steinacker:2016vgf, Sperling:2017dts, Sperling:2019xar, Steinacker:2019fcb, Fredenhagen:2021bnw, Steinacker:2022yhs, Steinacker:2022jjv} (see also~\cite{Valenzuela:2017ymx}).

In order to construct a sensible notion of distance over discrete parameter spaces, we introduce \emph{multicritical} vector models and discuss a possible bulk interpretation including Chan-Paton dressings of the usual higher-spin duality. Strikingly, we find that the higher-spin points do lie at infinite information distance, but the anomalous dimensions of higher-spin currents do not vanish exponentially, rather in a power-like fashion. This departure from the expected stringy behavior may be ascribed to the difference between vector and matrix models, as we shall explain in detail and corroborate with examples. Whether this indicates an inconsistency of these models in the sense of the swampland, or merely a separation from the string landscape, is unclear. To further investigate this issue, we explore the theory space in the direction spanned by the Chern-Simons coupling. It turns out that, consistently with the above observations, the decay of higher-spin anomalous dimensions is exponential along this direction, where indeed matrix-like gauge degrees of freedom dominate, at least in the regime that we study.

The paper is structured as follows. In~\cref{sec:SDC} we review the swampland distance conjecture, focusing on its CFT variant, and we describe its extension involving the quantum information metric. In~\cref{sec:bose_models} we introduce multicritical bosonic vector models, computing large-$N$ beta functions and the resulting RG flows in~\cref{sec:bose_beta_functions}, while focusing on the information metric in~\cref{sec:bosonic_metric_large-N,sec:bosonic_metric_eps-exp}. In~\cref{sec:fermi_models} we mirror the analysis for the fermionic case. In~\cref{sec:distances} we collect our results and use them to compute information distances for the models at hand, studying higher-spin limits to uncover their infinite-distance properties. In~\cref{sec:bosonic_dist_results,sec:fermionic_dist_results}, we present the distances for the bosonic and fermionic models respectively, while in~\cref{sec:cs_matter} we introduce a Chern-Simons sector in order to study limits in which matrix-like degrees of freedom dominate over vector-like ones in the information metric. In~\cref{sec:HS_holography} we recall the basics of higher spin holography in order to conclude
in~\cref{sec:swampland_implications} with some closing comments on the implications for the swampland program, higher-spin gravity and its separation -- if any -- from the string landscape. In~\cref{sec:gff} we discuss the unrelated case of generalized free fields as an example of theory space where a higher-spin limit does not lie at infinite information distance, and we propose a rationale for this result.

\section{Distance conjecture for CFTs and beyond}\label{sec:SDC}

\begin{figure}[ht!]
    \centering
    \includegraphics[scale=0.5]{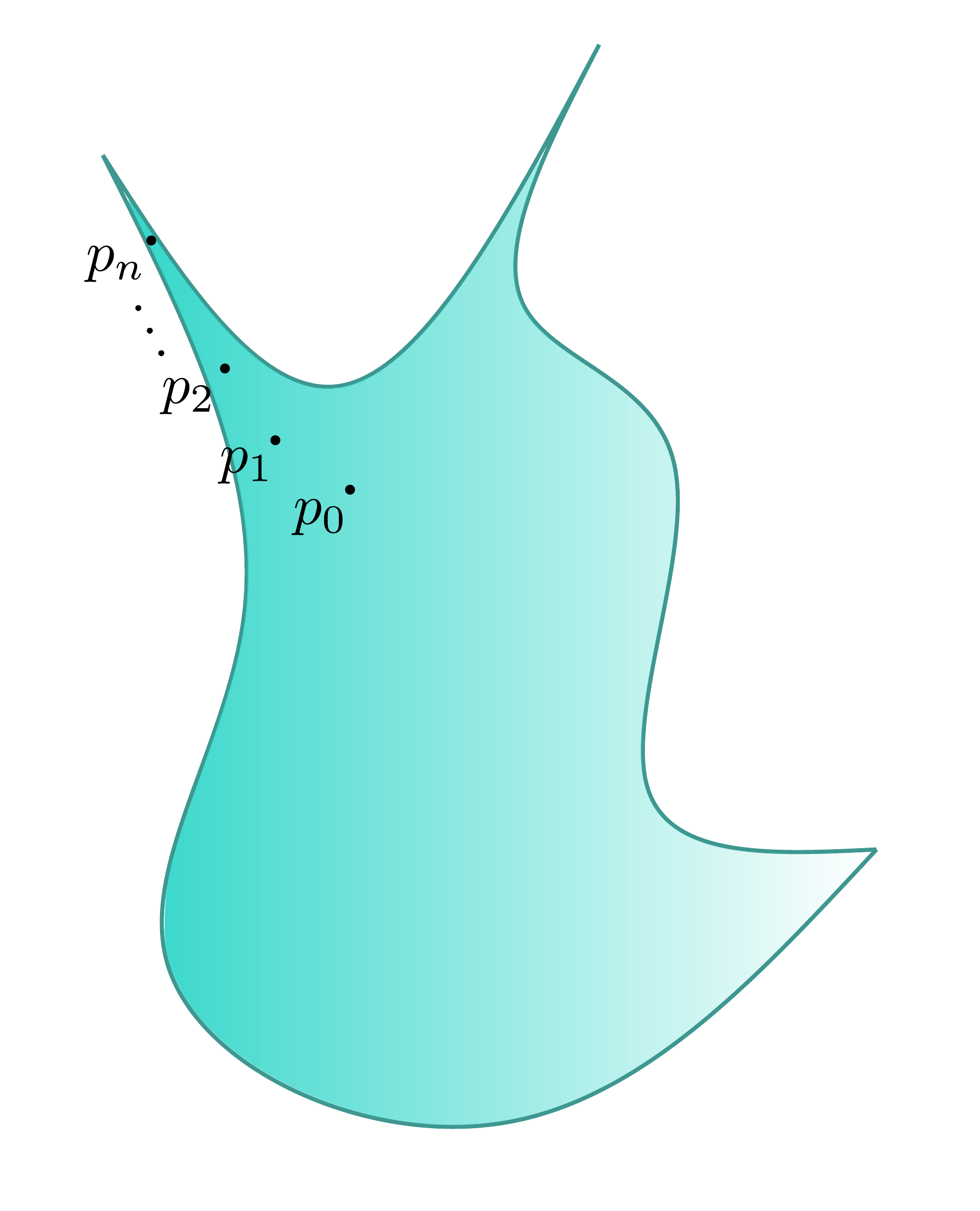}
    \caption{A sketch of a generic moduli space with infinite-distance points. They typically correspond to weakly coupled regimes (perhaps in a dual frame), thus we start the sequence $(p_n)_{n \geq 0}$ with $p_0$ already close to an infinite-distance point, and approach it as $n \to \infty$.}
    \label{fig:moduli_space}
\end{figure}

As we briefly recalled in the introduction, the swampland program aims at identifying general principles to distinguish those theories of quantum gravity that can be consistently completed in the UV from the ones that cannot. These proposed principles form a deeply interconnected web of ideas in which almost every conjecture exists independently of any particular top-down construction. The distance(/duality) conjecture~\cite{Ooguri:2006in} stands on somewhat different grounds: while it bears connections to other theory-independent swampland constraints, its underlying rationale seems to be more closely connected to string theory proper rather than a model-independent quantum gravity principle, in particular because of string dualities. For the purposes of this paper, the two key claims of the distance conjecture are the following: approaching an infinite-distance $\Delta \to \infty$ limit in moduli space, asymptotically
\begin{enumerate}
    \item an infinite tower of states becomes massless, and
    \item the masses $m$ vanish \emph{exponentially} fast along a sequence $(p_n)_{n \geq 0}$ of points in moduli space, according to\footnote{Here, and in the following, we denote by $\mathcal{O}(1)$ any positive constant that does not diverge nor vanish in the asymptotic regime at stake. Bounding the decay rate remains an important problem, but some partial results have been achieved~\cite{Andriot:2020lea, Baume:2016psm, Perlmutter:2020buo, Etheredge:2022opl}.}
    \begin{eqaed}\label{eq:SDC}
        \frac{m(p_n)}{m(p_0)} \overset{n \to \infty}{\sim} \mathcal{O}(1) \, e^{- \mathcal{O}(1) \, \Delta(p_n,p_0)} \, .
    \end{eqaed}
\end{enumerate}
The clear distinction between these two points will turn out to be crucial for our considerations. The latter statement can be equivalently phrased in terms of geodesics in moduli space, but the formulation above, employed in~\cite{Perlmutter:2020buo} and depicted in~\cref{fig:moduli_space}, is more general and more convenient for our purposes, since it more easily lends itself to extensions (see also~\cite{Lanza:2022zyg} for a discussion on different ways to approach infinite-distance points).

The distance conjecture has been refined and generalized in many directions~\cite{Baume:2016psm,Klaewer:2016kiy,Lee:2018urn, Lee:2019xtm, Lee:2019wij, Lust:2019zwm,Andriot:2020lea, Calderon-Infante:2020dhm, Baume:2020dqd, Perlmutter:2020buo, Lanza:2020qmt, Stout:2021ubb, Lanza:2021udy, Lanza:2022zyg, Grimm:2022sbl, Stout:2022phm}, but in this paper we will concern ourselves with two of them in particular. The first is the holographic, or CFT version of the conjecture~\cite{Baume:2020dqd, Perlmutter:2020buo}. It can be formulated very precisely for CFTs admitting a conformal manifold, where the tower of states is dual to a higher-spin tower of currents and the masses are replaced by their dual anomalous dimensions $\gamma_{\text{HS}}$. In this context, the moduli space is identified with the conformal manifold of the CFT, and the relevant metric that defines distances is the Zamolodchikov metric. The holographic dictionary then allows one to recover the bulk distance dividing by $\sqrt{C_T}$, the square-root of the central charge defined by the stress-tensor correlators~\cite{Baume:2020dqd}. The CFT distance conjecture of~\cite{Baume:2020dqd, Perlmutter:2020buo} concerns conformal manifolds, therefore the additional factor of $\sqrt{C_T}$ is immaterial, since it remains constant on the manifold.

However, we are interested in generalizing these ideas to settings where a conformal manifold is absent. A natural extension of the Zamolodchikov metric to the theory space of quantum field theories is the \emph{(quantum) information metric}~\cite{Provost:1980nc, Wootters:1981ki, oconnor1993geometry, Dolan:1997cx}, which was recently revisited in~\cite{Stout:2021ubb, Basile:2022zee, Stout:2022phm} in the swampland context. More specifically, the notion of information metric is both quite general and unique. Extracting a metric from a (family of) probability distribution(s) is essentially unique, but the choice of the distribution leaves at least two seemingly inequivalent choices: the distribution described by the ground states of a family of theories~\cite{Stout:2021ubb} or the one described by a (Wick-rotated) functional integral of a family of theories, namely the free energy~\cite{oconnor1993geometry, Dolan:1997cx} (see also~\cite{Anselmi:2011bp} for a related notion of metric). In this paper we consider the latter for a few reasons, namely its manifest spacetime covariance and the ease of computability.

The information metric as originally defined in~\cite{oconnor1993geometry} employs the family of distributions $e^{-S}/Z = e^{W-S}$ in field space, where the classical (Euclidean) action $S$ depends on some parameters (\emph{e.g.} couplings) $\{\lambda_i\}$. Thus, also the (source-free) partition function $Z$ and the free energy $W$ depend on the $\lambda_i$. The metric is defined by the vacuum expectation value (VEV)
\begin{eqaed}\label{eq:information_metric_def}
    G = \langle (dW-dS) \otimes (dW-dS) \rangle \, ,
\end{eqaed}
and its tensor components $G_{ij}(\lambda)$ transform accordingly under coordinate changes in theory space. If the action $S$ is \emph{linear} in the $\lambda_i$, one finds the more manageable expression
\begin{eqaed}\label{eq:information_metric_linear}
    G_{ij} = - \, \partial_i \partial_j w
\end{eqaed}
in terms of the ``Euclidean density'' of free energy $w = - \, \frac{1}{\text{Vol}} \, \log Z$, where $\text{Vol}$ is the Euclidean spacetime volume. In practice, this is the formulation we shall use to compute information distances.

Extending the distance conjecture to settings where exact moduli spaces or conformal manifolds are unavailable is interesting in its own right, as well as crucially important to study non-supersymmetric models~\cite{Basile:2022zee}. For the purposes of the present paper, our underlying motivation is to study infinite-distance limits in the context of higher-spin gravity. We restrict ourselves to the conjectural holographic duals of vector models~\cite{Klebanov:2002ja, Sezgin:2003pt, Leigh:2003gk, Giombi:2011kc, Bekaert:2012ux, Giombi:2012ms, Giombi:2016ejx}. A useful feature of the higher-spin/vector models duality is the zoo of non-supersymmetric models of this type. The parameter that governs the anomalous dimensions of (weakly broken) higher-spin currents is the rank $N \gg 1$ of the symmetry group~\cite{Giombi:2011kc,Giombi:2012ms,Skvortsov:2015pea, Giombi:2016hkj, Manashov:2016uam, Manashov:2017xtt, Gerasimenko:2021sxj}. In order to assess whether it makes sense to talk about infinite-distance limits in higher-spin gravity and, if so, what is its status in relation to the swampland, it is thus natural to employ the information metric of~\cite{oconnor1993geometry, Dolan:1997cx} to the theory space of vector models, connecting different symmetry groups via RG flows.

To this end, the simplest setting for our construction is a slight generalization of ordinary critical vector models: in this paper we examine RG flows and information metrics in \emph{multicritical} models~\cite{PhysRevB.13.412, Eyal:1996da, Pelissetto:2000ek, Pelissetto:2001fi, Calabrese:2002bm, folk2009field, Eichhorn:2013zza, Rychkov:2018vya, Vacca:2019rsh, Chai:2020hnu, Chai:2020zgq}, where the symmetry group is extended to a product of factors. The upshot of this intuitive generalization of the (CFT) distance conjecture is a concrete result regarding the two basic tenets described above. Specifically, we find that the large-$N$ limits in which higher-spin symmetry is recovered indeed lie at infinite distance. However, as we shall discuss in detail in~\cref{sec:distances}, it turns out that the anomalous dimensions of higher-spin currents appear not to decay exponentially, but rather polynomially fast. \emph{A posteriori} this striking departure from the expected behavior is not especially surprising, since the exponential behavior seems characteristic of strings. Furthermore, it is not clear whether a sensible finite-rank bulk description of higher-spin theories of gravity exists, although settings such as ABJ triality~\cite{Chang:2012kt} may help shed light on this matter. As we shall discuss in detail in~\cref{sec:cs_matter}, already the coupling space of Chern-Simons-matter CFTs exhibits stringy matrix-like degrees of freedom, and does lead to exponential decay in this direction of theory space.

In the following sections we present the multicritical field theories at stake, studying RG flows and computing the information distance along suitable trajectories in theory space.

\section{Multicritical bosonic models}\label{sec:bose_models}

In this section we describe in detail the bosonic models that we consider. The well-known critical $O(N)$ vector models can be generalized to multicritical models, whose field content comprises spacetime scalars
\begin{eqaed}\label{eq:multicritical_content}
    \phi_1 \, , \, \dots \, , \, \phi_k
\end{eqaed}
where for each $a = 1 \, , \, \dots \, , \, k$ the field $\phi_a$ belongs to the vector representation of $O(N_a)$. The symmetry group of the theory is thus $O(N_1) \times \dots \times O(N_k)$. In particular, the bicritical case $k=2$ plays a role in critical phenomena of condensed-matter systems~\cite{Calabrese:2002bm, Chai:2020hnu}.\footnote{The ``frustrated'' models of~\cite{Pelissetto:2001fi} are somewhat similar, but their theory space and RG flows are different.}

The quartic couplings that preserve this symmetry are packaged into a symmetric, positive-definite matrix $\lambda_{ab}$, and the Euclidean action reads
\begin{eqaed}\label{eq:multicritical_action}
    S = \int d^dy \left( \frac{1}{2} \, (\partial \mathbf{\phi}_a)^2 + \frac{1}{2} \, r_a \, \mathbf{\phi}_a^2 + \frac{\lambda_{ab}}{N} \, \mathbf{\phi}_a^2 \, \mathbf{\phi}_b^2 \right) .
\end{eqaed}
In order to simplify the ensuing analysis, let us introduce some notation. We write $N = \sum_a N_a$ for the total rank and we define the ratios $x_a = \frac{N_a}{N}$. For $d$ spacetime dimensions, one can further define dimensionless couplings $r_a = \mu^2 \, g_a$ and $\lambda_{ab} = \mu^{4-d} \, g_{ab}$ in terms of the RG scale $\mu$.
 
\subsection{Bosonic beta functions}\label{sec:bose_beta_functions}

In order to study infinite-distance limits of RG flows in multicritical models we shall focus on two limiting regimes: the $\epsilon$-expansion and the large-$N$ limit. The latter is defined by $N \gg 1$ with $x_a$ and $g_{ab}$ fixed, both for the bosonic and the fermionic case. The former is defined by $\epsilon \ll 1$ with $d = 4 - \epsilon$ in the bosonic case and $d = 2 + \epsilon$ in the fermionic case. Since we are interested in infinite-distance limits, we work at leading order in the large-$N$ limit in the remainder of this paper, unless otherwise stated. Since this is not sufficient to perform all computations in closed form, the $\epsilon$-expansion will serve as an additional analytical tool to obtain explicit expressions. We will then compare the results with the numerics for $\epsilon = 1$.

\begin{table}[ht!]
    \centering
    \begin{tabular}{r|c|c}
        & $\epsilon = 1$ & $\epsilon \ll 1$ \\
        \hline
        Bose ($d = 4-\epsilon$) & large $N$ (numeric) & pert. exp. $N \gg 1$ (analytic) \\
        \hline
        Fermi ($d = 2 + \epsilon$) & large $N$ (numeric) & pert. exp. $1/\epsilon \gg N \gg 1$ (analytic)
    \end{tabular}
    \caption{Our scheme of approximations.}
    \label{tab:approximations}
\end{table}

The first step toward infinite-distance limits is to compute beta functions. Both in the $\epsilon$-expansion and at large $N$ the beta functions are dominated by the one-loop contribution. In the bosonic model, the one-loop beta functions were worked out in~\cite{Rychkov:2018vya, Chai:2020hnu}, which we recovered via a heat-kernel computation. The resulting beta functions read\footnote{Here, and in the following, we leave summations over $a, b, c, d$ dummy indices implicit, unless there is a potential ambiguity.}
\begin{eqaed}\label{eq:eps-exp_rge}
    \mu\, \frac{dg_a}{d\mu} & = - \, 2 g_a - 2B \left(x_b - \, \frac{4B}{N} \, \delta_{ab} \right) \frac{g_{ab}}{1 + g_a} \, , \\
    \mu\, \frac{dg_{ab}}{d\mu} & = - \, \epsilon \, g_{ab} + 2B \left(x_c + \frac{2}{N} \left( \delta_{ac} + \delta_{bc} \right) \right) \frac{g_{ac} \, g_{bc}}{(1 + g_c)^2} + \frac{8B}{N} \, \frac{g_{ab}^2}{(1 + g_a)^2 (1 + g_b)^2} \, ,
\end{eqaed}
where $B \equiv \frac{2\Omega_{d-1}}{(2\pi)^d} = \frac{1}{4\pi^2} + \mathcal{O}(\epsilon)$ and $\Omega_{d-1}$ is the volume of the $(d-1)$-dimensional unit sphere. Let us remark that, for large $N$,~\cref{eq:eps-exp_rge} are reliable even for $\epsilon = 1$, namely in $d=3$.

Since we are going to study RG flows connecting critical points, we work in a massless scheme. To further simplify the flow equations, we define rescaled couplings as in~\cite{Chai:2020hnu}:
\begin{eqaed}\label{eq:rescaled_couplings}
    g_{ab} \equiv \frac{4-d}{2B} \, \alpha_{ab} \overset{\epsilon \ll 1}{\sim} 2\pi^2 \epsilon \, \alpha_{ab} \, ,
\end{eqaed}
as well as the RG time
\begin{eqaed}\label{eq:RG_time}
    t \equiv (4-d) \log \frac{\mu}{\Lambda_\text{UV}}
\end{eqaed}
which absorbs the dependence on the UV cutoff in the RG flow. With these conventions,~\cref{eq:eps-exp_rge} can be recast in the simpler form
\begin{eqaed}\label{eq:eps-exp_rge_small_mass}
    \dot{\alpha}_{ab} & = - \, \alpha_{ab} + \left(x_c + \frac{2}{N} \left( \delta_{ac} + \delta_{bc} \right) \right) \alpha_{ac} \, \alpha_{bc} + \frac{4}{N} \, \alpha_{ab}^2 \, ,
\end{eqaed}
where the dot denotes a derivative with respect to the RG time $t$.

\subsubsection{Bosonic renormalization group flows}\label{sec:bosonic_RG_flows}

The RG flows described by~\cref{eq:eps-exp_rge_small_mass} possess many fixed points where the global symmetry of the theory in~\cref{eq:multicritical_action} is enhanced.\footnote{There also exist ``conical'' fixed points where different $O(N_a)$ are coupled but no symmetry enhancement occurs~\cite{PhysRevB.13.412, Calabrese:2002bm, Chai:2020hnu}.} This occurs whenever the coupling matrix $\alpha$ comprises decoupled diagonal blocks, each of which has the appropriate fixed-point value. For instance, the upper-left $m \times m$ block can decouple and acquire the fixed-point value corresponding to the critical $O\left(\sum_{a=1}^m N_a\right)$ model. In particular, there are flows where two factors $O(N_a)$ ``fuse'' according to
\begin{eqaed}\label{eq:bicritical_flow_symmetry}
    O(N_a) \times O(N_b) \quad \longrightarrow \quad O(N_a+N_b) \, ,
\end{eqaed}
while all other couplings stay fixed. The beta functions decouple and the RG flow is effectively bicritical.\footnote{Let us remark that, since the 't Hooft couplings $\alpha$ refer to the total rank $N$, the fixed-point values are rescaled with respect to the ordinary bicritical model.} For reasons that will become apparent in~\cref{sec:distances}, we will consider chains of RG flows composed of elementary steps of this type. This allows us to effectively restrict computations to simple bicritical models, up to some adjustments of the ranks and fixed-point values of the couplings. In this context one can thus display bicritical couplings as symmetric $2 \times 2$ matrices. 

As we will explain in more detail in~\cref{sec:distances}, we will focus on $O(N)^M$ multicritical models, where all the ranks are equal and the chain of RG flows comprises $M-1$ steps. Specifically, it takes the form 
\begin{eqaed}\label{eq:bosonic_bicritical_flow}
   O(MN) & \quad \longrightarrow \quad O((M-1)N) \times O(N) \quad \longrightarrow \quad O((M-2)N) \times O(N)^2 \\
   & \qquad \longrightarrow \quad \dots \quad \longrightarrow \quad O(N) \times O(N)^{M-1} \quad \longrightarrow \quad O(N)^M
\end{eqaed}
in the bosonic case, while in the fermionic case the flow goes in the opposite direction. The $k$-th step of the chain is
\begin{eqaed}\label{eq:bose_kth_step}
    & O((k+1)N) \times O(N)^{M-k-1} \quad \longrightarrow \quad O(kN) \times O(N)^{M-k} \, .
\end{eqaed}
At each step, we dub the corresponding fixed points FP$_S$ and FP$_B$, since the symmetry group $O((k+1)N)$ at the ``symmetric'' point FP$_S$ goes to $O(kN) \times O(N)$ at the ``broken'' point FP$_B$. The matrices of couplings $\alpha \equiv \frac{2B}{4-d} \, \mathbf{g}$ are dimension-independent at the fixed points, where they take the values
\begin{eqaed}\label{eq:bosonic_FPS_FPB}
    \text{FP}_S \, : \, \mqty(\frac{MN}{(k+1)N+8} & \frac{MN}{(k+1)N+8} \\ \frac{MN}{(k+1)N+8} & \frac{MN}{(k+1)N+8}) \quad \longrightarrow \quad \text{FP}_B \, : \, \mqty(\frac{MN}{kN+8} & 0 \\ 0 & \frac{MN}{N+8}) \, .
\end{eqaed}

\begin{figure}[ht!]
    \centering
    \includegraphics[scale=0.7]{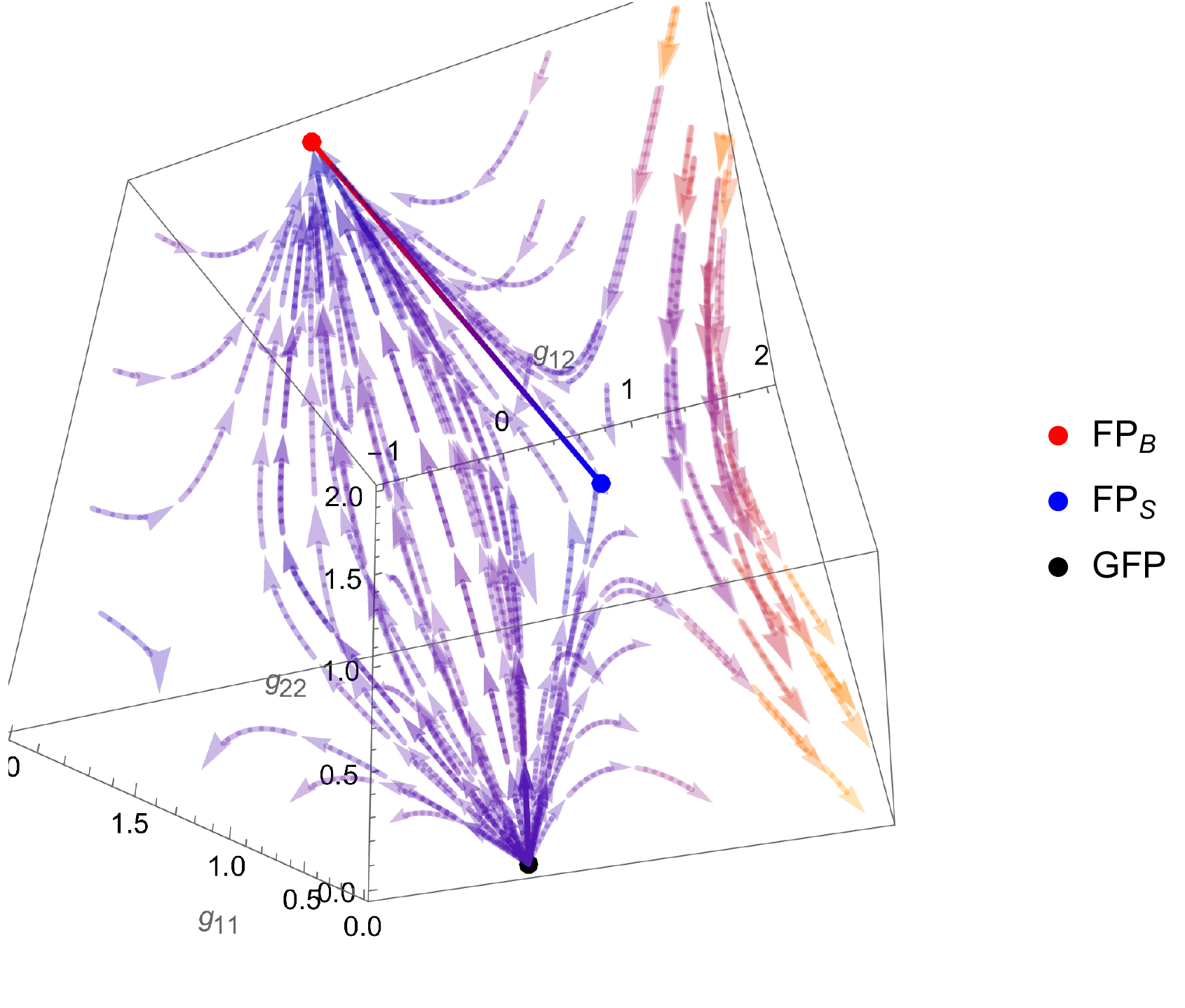}
    \caption{Streamplot of the RG flow for the bosonic model for the bicritical flow with both ranks equal to 1000. The straight line represents a single step of the type of~\cref{eq:bose_kth_step}. The direction of the flow is following the directions of the arrows, or equivalently from blue to red on the highlighted single-step trajectory. The fixed points FP$_S$ (resp. FP$_B$) corresponds to the fixed point of the (un)broken configuration, while GFP represents the Gaussian fixed point of zero couplings.}
    \label{fig:bose_streamplot}
\end{figure}

The flow diagram, restricted to the bicritical subspace, is depicted in~\cref{fig:bose_streamplot}. Examining the beta functions of~\cref{eq:eps-exp_rge_small_mass}, one finds that at large $N$ the specific trajectory of~\cref{eq:bosonic_FPS_FPB} is simply the line connecting FP$_S$ to FP$_B$. Since the information distance is reparametrization invariant, we choose to parametrize it with $u \in (0,1)$ according to
\begin{eqaed}\label{eq:bose_RG_line}
    \alpha(u) & = \alpha_{\text{FP}_S} \, (1-u) + \alpha_{\text{FP}_B} \, u \\
    & \equiv \alpha_{\text{FP}_S} + v \, u \, ,
\end{eqaed}
where $v \equiv \alpha_{\text{FP}_B} - \alpha_{\text{FP}_S}$ is the tangent vector to the line.

\subsection{Information metric at large \texorpdfstring{$N$}{N}}\label{sec:bosonic_metric_large-N}

In this section we set up the computation of the information metric in the bosonic model. Since we are interested in infinite-distance limits, we work at large $N$.

We begin introducing Hubbard-Stratonovich fields $\sigma_a$, to rewrite the quartic interactions in terms of the cubic interactions $\sigma_a \phi_a^2$. Integrating out the $\phi_a$, the resulting effective action takes the form~\cite{Eyal:1996da, Chai:2020hnu}\footnote{The same result can be obtained with a variational calculation~\cite{Chai:2020zgq}.}
\begin{eqaed}\label{eq:hubbard_stratonovich_action}
    S_\text{eff} = - \, \frac{N}{4} \, (g^{-1})^{ab} \, \int d^dy \, \sigma_a \, \sigma_b + \frac{N}{2} \, x_a \, \text{Tr} \log \left(- \, \Box + r_a + 2 \, \sigma_a \right) ,
\end{eqaed}
where $g^{-1}$ denotes the matrix inverse of $g$. The above expression is now amenable to a saddle-point large-$N$ expansion, where the $x_a$ are kept fixed.\footnote{In principle one could also build a large-$N$ expansion where some of the $N_a = N x_a$ are fixed instead~\cite{Chai:2020hnu, Chai:2020zgq}.}

For constant $\sigma_a$, the effective potential reads
\begin{eqaed}\label{eq:eff_potential}
    V_\text{eff} = - \, \frac{N}{4} \, (g^{-1})^{ab} \, \sigma_a \, \sigma_b + \frac{N}{2} \, x_a \int^{\Lambda_\text{UV}} \frac{d^dp}{(2\pi)^d} \log\left(p^2 + r_a + 2 \, \sigma_a \right) ,
\end{eqaed}
and the saddle-point condition yields the gap equations
\begin{eqaed}\label{eq:gap_eqs}
    \sigma_a = 2 \, g_{ab} \, x_b \, \dot{L}(r_b + 2 \sigma_b) \, ,
\end{eqaed}
where $\dot{L}(r) \equiv \frac{dL}{dr}$ and we have defined
\begin{eqaed}\label{eq:log_propagator}
    L(r) & \equiv \int^{\Lambda_{\text{UV}}}\frac{d^dp}{(2\pi)^d} \, \log(p^2 + r) \\
    & = \frac{\Omega_{d-1}}{d(2\pi)^d} \, \frac{\Lambda_{\text{UV}}^d}{r} \, _2F_1\left(1, \frac{d}{2} ; \frac{d+2}{2} ; - \, \frac{\Lambda_{\text{UV}}^2}{r}\right) .
\end{eqaed}
Finally, denoting the saddle-point gaps by $\overline{\sigma}_a(\mathbf{r}, \mathbf{g})$ as a function of the couplings, the large-$N$ negative energy density $- \, w = \frac{1}{\text{Vol}} \, \log Z$ takes the form
\begin{eqaed}\label{eq:large-N_energy_density}
    - \, w \overset{N \gg 1}{\sim} - \, V_\text{eff}(\overline{\sigma}(\mathbf{r}, \mathbf{g})) \, ,
\end{eqaed}
and its Hessian in coupling space
\begin{eqaed}\label{eq:large-N_metric}
    G_{(ab)(cd)} = - \, \frac{\partial}{\partial g_{ab}} \frac{\partial}{\partial g_{cd}} V_\text{eff}(\overline{\sigma}(\mathbf{0}, \mathbf{g}))
\end{eqaed}
gives the (intensive~\cite{Stout:2021ubb}) quantum information metric restricted to the subspace of vanishing quadratic couplings. While ordinary diagrammatic perturbation theory in this subspace is subject to IR divergences, we expect the non-perturbative result to be well-defined. Indeed, we have verified that the analytic result from the $\epsilon$-expansion matches the numerical computation of~\cref{eq:large-N_metric}. Furthermore, we have verified that, in the ordinary $O(N)$ model,~\cref{eq:large-N_metric} matches the calculation of~\cite{Dolan:1997cx}, including all the combinatoric factors. 

\subsection{Information metric in the \texorpdfstring{$\epsilon$}{Epsilon}-expansion}\label{sec:bosonic_metric_eps-exp}

Let us now compute the information metric in the $\epsilon$-expansion. We are interested in critical RG trajectories which remain massless, but we first present the result in the massive case since it is simpler. We will then extend our technique to the massless case.

Since for $\epsilon \ll 1$ the complete RG flow along a critical trajectory remains perturbative, the information metric can be computed from the second-order\footnote{The first-order term can be computed more easily. However, it is irrelevant when evaluated on critical RG trajectories in a massless scheme, since the reduced metric involves two derivatives with respect to the quartic couplings only.} (negative) vacuum energy density
\begin{eqaed}\label{eq:second-order_logZ_multicritical}
   - \, w_2(\mathbf{g}) & \equiv \frac{1}{\text{Vol}} \, \log Z|_{\mathcal{O}(\mathbf{g}^2)} \, ,
\end{eqaed}
which is calculable in perturbation theory. Naively, a diagrammatic computation would yield
\begin{eqaed}\label{eq:second-order_logZ_diagrams}
    - \, w_2(\mathbf{g}) & = \frac{4\left(g_{ab}^2 N_a N_b + 2 N_a g_{aa}^2\right)}{(2\pi)^{3d} N^2} \, \mathcal{I}_{\text{banana}} \\ & + \frac{4 g_{ab} g_{bc} N_a N_b N_c + 16 g_{aa} g_{ab} N_a N_b + 16 g_{aa}^2 N_a}{(2\pi)^{3d} N^2} \, \mathcal{I}_{\text{bubble}} \, ,
\end{eqaed}
where the (UV regulated) integrals stemming from the banana and triple-bubble Feynman diagrams would be
\begin{eqaed}\label{eq:sunset_banana}
    \mathcal{I}_{\text{banana}} & \equiv (2\pi)^{3d} \int d^dy \, G_0(y)^4 = \int \frac{d^dp \, d^dq \, d^dk}{p^2 q^2 k^2 (p+q+k)^2} \, , \\
    \mathcal{I}_{\text{bubble}} & \equiv (2\pi)^{3d} \int d^dy \, G_0(0)^2 \, G_0(y)^2 = \int \frac{d^dp \, d^dq \, d^dk}{p^2 q^2 k^4} \, .
\end{eqaed}
However, it is clear that in the massless case these expressions are IR-divergent. Since we shall work in the large $N$, both for $\epsilon = 1$ and $\epsilon \ll 1$, the appropriate treatment that avoids these IR issues involves solving~\cref{eq:gap_eqs} perturbatively at weak coupling. As a result, the correct asymptotics of the loop integrals as functions of the gaps $\sigma_a$ includes a logarithmic term. Indeed, one finds
\begin{eqaed}\label{eq:bose_eps-exp_gap_sol}
    \overline{\sigma}_a \overset{\epsilon \ll 1}{\sim} \frac{B}{2} \, g_{ab} \, x_b \, ,
\end{eqaed}
so that the relevant terms in the (negative) free energy (density), corresponding to the leading-order terms in the metric, read
\begin{eqaed}\label{eq:pert_free_energy_r=0}
    - \, w_2(\mathbf{g}) & \overset{\epsilon \ll 1}{\sim} \frac{N}{16} \, B^3 \, x_a \, x_b \, x_c \, g_{ab} \, g_{bc}  \left( \frac{1}{2}  - \, \log \left(B \, g_{bd} \, x_d\right) \right) ,
\end{eqaed}
in units of the UV cutoff. The resulting metric can be readily computed via~\cref{eq:large-N_metric}, taking special care of the fact that the matrix derivatives are to be taken along the subspace of symmetric matrices.

\section{Multicritical fermionic models}\label{sec:fermi_models}

Let us now describe the fermionic counterparts of the models that we discussed thus far. The generalization now involves the Gross-Neveu model. Its multicritical version features fermions $(\psi_a)_{a=1\,,\,\dots\,,\,k}$, each of which belongs to the defining representation of $U(N_a)$, so that the symmetry group is $U(N_1) \times \dots \times U(N_k)$. By and large, the treatment in this section closely follows the one for the bosonic case, hence we will highlight the main differences as we proceed.

To begin with, in the presence of spinors, Wick rotation to Euclidean signature needs slightly more care. Our starting point is thus the Lorentzian (mostly-plus) action
\begin{eqaed}\label{eq:multicritical_gn}
    S = \int d^dy \left(- \, \overline{\psi}_a \slashed{\partial} \psi_a + \frac{\lambda_{ab}}{N} \, (\overline{\psi}_a \psi_a)(\overline{\psi}_b \psi_b) \right) ,
\end{eqaed}
and we introduce Hubbard-Stratonovich singlet fields $\sigma_a$ analogously to the bosonic case. The resulting auxiliary action is
\begin{eqaed}\label{eq:multicritical_gn_aux}
    S = \int d^dy \left( - \, \overline{\psi}_a \slashed{\partial} \psi_a + \sigma_a (\overline{\psi}_a \psi_a) - \, \frac{N \lambda^{-1}_{ab}}{4} \, \sigma_a \sigma_b \right) .
\end{eqaed}
One can then perform a Wick rotation, which replaces $- \slashed{\partial}_L \to \slashed{\partial}_E$ and changes the sign of the other terms. 

\subsection{Fermionic beta functions}\label{sec:fermi_beta_functions}

Following the computation in the preceding section, one would integrate out the fermions and seek saddle points with constant gaps $\sigma_a$. The renormalization of the couplings due to UV-divergent terms in the functional determinant contain the leading-order terms in the beta function. However, in order to match the next-to-leading order accuracy of~\cref{eq:eps-exp_rge_small_mass}, one needs to integrate out the fermions around a non-vanishing background. Performing a general background-field expansion to second order in the fluctuations $\delta \sigma_a \, , \, \delta \psi_a \, , \, \delta \overline{\psi}_a$ around a constant background, the quadratic kinetic operator for fluctuations takes the form
\begin{eqaed}\label{eq:kinetic_operator_gn}
    \mathcal{D} = \mqty( 0 & \slashed{\partial} - \sigma_a & \psi_a \\
    \slashed{\partial} - \sigma_a & 0 & \overline{\psi}_a \\
    \psi_a &  \overline{\psi}_a & \frac{N \lambda^{-1}_{ab}}{2}) \, .
\end{eqaed}
Setting the fermionic background to zero yields the leading-order large-$N$ result for the effective potential. Since the kinetic matrix mixes bosonic and fermionic fluctuations, for complex bosons the formula
\begin{eqaed}\label{eq:block_det}
    \det \mqty(\rm A & \rm B \\ \rm C & \rm D) = \det \rm A \, \det(\rm D - C \, A^{-1} \, B) \, ,
\end{eqaed}
for block matrices is replaced by the Berezinian\footnote{The conventions of~\cref{eq:kinetic_operator_gn} translate into~\cref{eq:block_det} written with the ``Fermi-Fermi'' block in the upper-left corner.}
\begin{eqaed}\label{eq:block_berezinian}
    \frac{\det(\rm D - C \, A^{-1} \, B)}{\det \rm A} \, .
\end{eqaed}
In our case, since the bosons are real, the bosonic contribution in the numerator comes with a square root, which translates into a factor of $\frac{1}{2}$ in the corresponding term in the effective action. One thus finds that, up to a constant irrelevant for the beta function and the effective potential, the contribution to the one-loop effective action is
\begin{eqaed}
    - \, N_a \, \Tr \log(-\Box + \sigma^2_a) + \frac{1}{2} \, \Tr \log(1 - \frac{4}{N} \, \mathbf{\lambda} \, \overline{\psi} ( \slashed{\partial} - \mathbf{\sigma})^{-1} \psi) \, ,
\end{eqaed}
where $\mathbf{\sigma} = \text{diag}(\sigma_a)$ and $\mathbf{\lambda} = (\lambda_{ab})$ are intended as $k \times k$ matrices in multicritical flavor space. In order to obtain the beta functions, one can expand the logarithms keeping the terms that correct the vertices. To this end, one can observe that the last term contributes
\begin{eqaed}\label{eq:last_gn_contribution}
    \frac{2}{N} \, \lambda_{aa} \, \sigma_a \overline{\psi}_a \psi_a \, I_2 \, , \qquad I_2 \equiv \int^{\Lambda_{\text{UV}}} \frac{d^dp}{(2\pi)^d} \frac{1}{p^2}
\end{eqaed}
while the first term contributes $- \, N_a \, \sigma_a^2 \, I_2$. The next step is to integrate out the $\sigma_a$ fields to obtain the corrected coupling, but one can simplify matters observing that the correction to the saddle-point value $\sigma_a = \frac{2}{N} \, \lambda_{ab} \overline{\psi}_b \psi_b$ does not contribute to one-loop order. Therefore, one finds the corrected Euclidean coupling
\begin{eqaed}\label{eq:corrected_coupling_gn}
    & - \, \frac{1}{N} \left(\lambda_R\right)_{ab} (\overline{\psi}_a \psi_a) (\overline{\psi}_b \psi_b) \, , \\
    & \left(\lambda_R\right)_{ab} = \lambda_{ab} + \frac{4}{N} \, I_2 \left(N_c \, \lambda_{ac} \lambda_{bc} - \frac{1}{2} \left(\lambda_{aa} + \lambda_{bb}\right) \lambda_{ab} \right) .
\end{eqaed}
From this expression one can extract the beta functions. Letting the dimensionless running couplings $g_{ab} \equiv \mu^{d-2} \, \lambda_{ab}$, one finds
\begin{eqaed}\label{beta_function_gn_dd}
    \beta_{ab}(\mathbf{g}) = (d-2) \, g_{ab} - \, \frac{2B}{N} \left(N_c \, g_{ac} \, g_{bc} - \frac{1}{2} \left(g_{aa} + g_{bb}\right) g_{ab} \right) ,
\end{eqaed}
and in particular to leading order in an $\epsilon$-expansion for $d = 2 + \epsilon$
\begin{eqaed}\label{eq:beta_function_gn_eps}
    \beta_{ab} = \epsilon \, g_{ab} - \, \frac{2}{\pi \, N} \left(N_c \, g_{ac} \, g_{bc} - \frac{1}{2} \left(g_{aa} + g_{bb}\right) g_{ab} \right) .
\end{eqaed}
In order to obtain an expression along the lines of~\cref{eq:eps-exp_rge_small_mass}, in analogy with~\cref{eq:rescaled_couplings} let us define
\begin{eqaed}\label{eq:fermi_rescaled_couplings}
    g_{ab} \equiv \frac{d-2}{2B} \, \alpha_{ab} \overset{\epsilon \ll 1}{\sim} \frac{\pi}{2} \, \epsilon \, \alpha_{ab} \, ,
\end{eqaed}
and once again $t \equiv (d-2) \log \frac{\mu}{\Lambda_\text{UV}}$. One finally arrives at a result similar to~\cref{eq:eps-exp_rge_small_mass},
\begin{eqaed}\label{eq:eps-exp_gn}
    \dot{\alpha}_{ab} = \alpha_{ab} - \left( x_c - \, \frac{1}{2N} \left(\delta_{ac} + \delta_{bc}\right) \right) \alpha_{ac} \, \alpha_{bc} \, .
\end{eqaed}
To our knowledge, the above is the first computation of the beta functions of the multicritical Gross-Neveu model in the literature.

\subsubsection{Fermionic renormalization group flows}\label{sec:fermi_RG_flows}

In the fermionic case, similarly to the bosonic one, we will consider models of the form $U(N)^M$ and compute the information distance as a function of $M \gg 1$. In this case, the chain of RG flows is
\begin{eqaed}\label{eq:fermionic_bicritical_flow}
   U(MN) & \quad \longleftarrow \quad U((M-1)N) \times U(N) \quad \longleftarrow \quad U((M-2)N) \times U(N)^2 \\
   & \quad \longleftarrow \quad \dots \quad \longleftarrow \quad U(N) \times U(N)^{M-1} \quad \longleftarrow \quad U(N)^M
\end{eqaed}
Analogously to~\cref{eq:bose_kth_step}, the single-step RG flow
\begin{eqaed}\label{eq:fermi_kth_step}
    U(kN) \times U(N)^{M-k} \quad \longrightarrow \quad U((k+1)N) \times U(N)^{M-k-1} \, .
\end{eqaed}
now connects the ``broken'' fixed point FP$_B$ to ``symmetric'' FP$_S$. The matrices of couplings $\alpha \equiv \frac{2B}{d-2} \, \mathbf{g}$ are once again independent of the dimension at the fixed points, where they take the values
\begin{eqaed}\label{eq:fermionic_FPB_FPS}
    \text{FP}_B \, : \, \mqty(\frac{MN}{kN-1} & 0 \\ 0 & \frac{MN}{N-1}) \quad \longrightarrow \quad \text{FP}_S \, : \, \mqty(\frac{MN}{(k+1)N-1} & \frac{MN}{(k+1)N-1} \\ \frac{MN}{(k+1)N-1} & \frac{MN}{(k+1)N-1}) \, .
\end{eqaed}

\begin{figure}[ht!]
    \centering
    \includegraphics[scale=0.7]{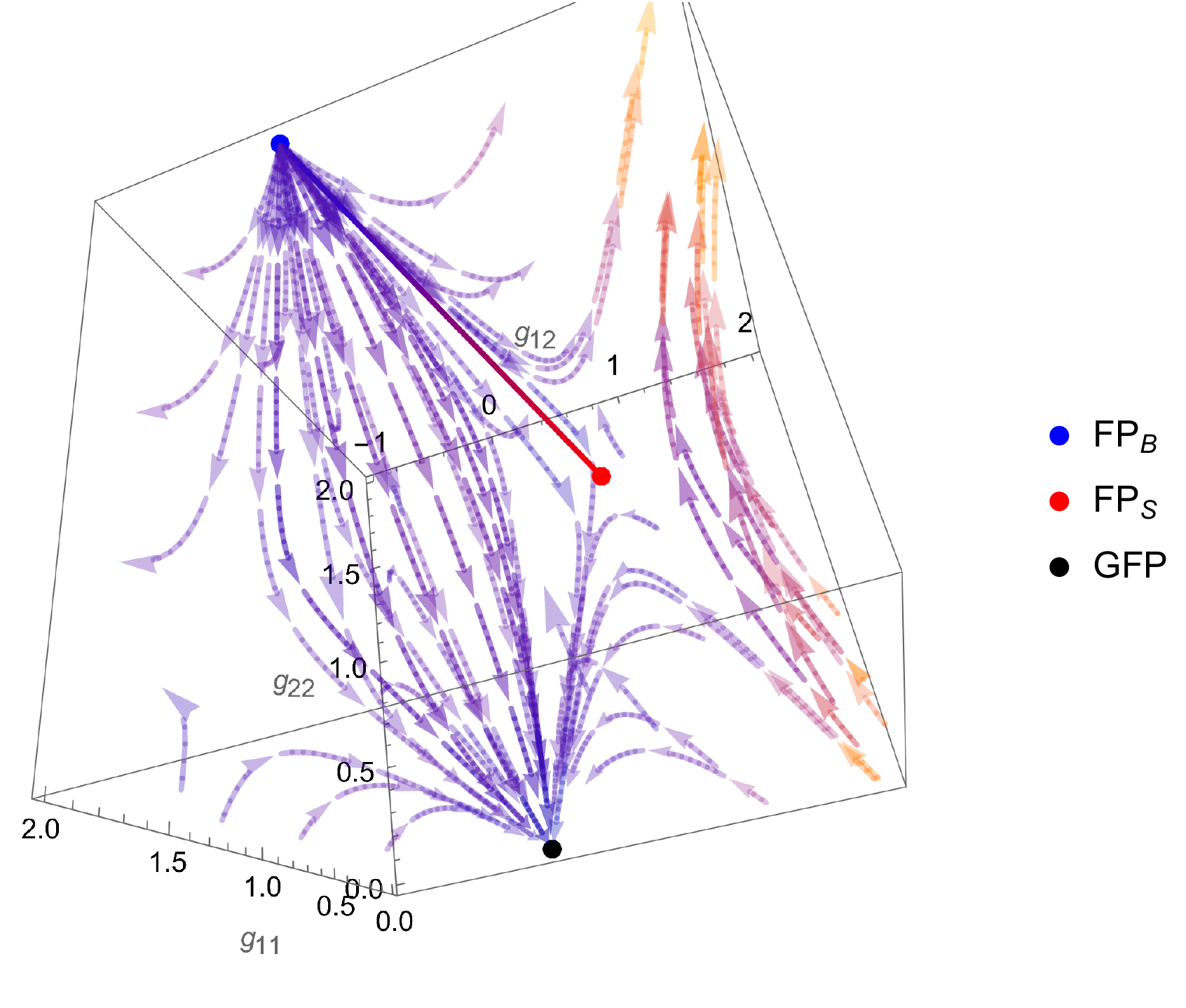}
    \caption{Streamplot of the RG flow for the fermionic model for the bicritical flow with both ranks equal to 1000. The straight line represents a single step of the type of~\cref{eq:fermi_kth_step}. The direction of the flow is following the directions of the arrows, or equivalently from blue to red on the highlighted single-step trajectory. The fixed points FP$_S$ (resp. FP$_B$) corresponds to the fixed point of the (un)broken configuration, while GFP represents the Gaussian fixed point of zero couplings.}
    \label{fig:fermi_streamplot}
\end{figure}

The flow diagram, restricted to the bicritical subspace, is depicted in~\cref{fig:fermi_streamplot}.

In complete analogy to the bosonic case, the relevant trajectory of~\cref{eq:fermionic_FPB_FPS} is simply a line to leading order in the large-$N$ limit. Thus, we choose to parametrize it according to
\begin{eqaed}\label{eq:fermi_RG_line}
    \alpha(u) & = \alpha_{\text{FP}_B} \, (1-u) + \alpha_{\text{FP}_S} \, u \\
    & \equiv \alpha_{\text{FP}_B} + v \, u \, ,
\end{eqaed}
where now $v \equiv \alpha_{\text{FP}_S} - \alpha_{\text{FP}_B}$ is properly oriented tangent vector to the line.

\subsection{Information metric for fermions}\label{sec:fermionic_metric}

Similarly to the bosonic case, evaluating the effective action on a background with constant gaps $\sigma_a$, one obtains the effective potential for the fermionic models. With the same definition in~\cref{eq:log_propagator}, it takes the form
\begin{eqaed}\label{eq:eff_potential_fermionic}
    V_\text{eff} = \frac{N}{4} \, (g^{-1})^{ab} \, \sigma_a \, \sigma_b - N \, x_a \, L(\sigma^2_a) \, ,
\end{eqaed}
and the corresponding saddle-point gap equations now take the form
\begin{eqaed}\label{eq:fermi_gap}
    \sigma_a = 4 \, g_{ab} \, x_b \, \sigma_b \, \dot{L}(\sigma^2_b) \, .
\end{eqaed}
In particular, for $d=2+\epsilon$, one has
\begin{eqaed}\label{eq:2d_log_propagator}
    \dot{L}(\sigma^2) \overset{\epsilon \ll 1}{\sim} \frac{B}{4} \, \log\left(1 + \frac{1}{\sigma^2}\right) .
\end{eqaed}

The full large-$N$ metric can then be computed numerically, but analogously to the bosonic case one expects the calculation to simplify dramatically for $\epsilon \ll 1$, since the entire RG trajectory is weakly coupled. However, since close to $d=2$ the Gross-Neveu model develops gaps that are exponentially suppressed in the couplings, one cannot apply conventional perturbation theory to approach~\cref{eq:fermi_gap}. Numerical precision is also compromised by exponential underflow. We were not able to solve~\cref{eq:fermi_gap} analytically for generic (small) couplings, but we managed to find a closed-form solution along the RG trajectory that we need to consider. Similarly, we managed to arrive at a closed-form expression for the pullback of the metric $G$ along the relevant RG trajectory. Although the general form of $G$ close to $d=2$ is unknown (and unnecessary for us), one can compute it numerically in $d=3$. Since we are ultimately interested in computing distances along a specific RG trajectory, the pullback method does not bring along any further complications.

Starting from~\cref{eq:large-N_metric}, one can explicitly write the derivatives via the chain rule, \emph{e.g.}
\begin{eqaed}\label{eq:metric_chain_rule}
    \frac{\partial V_{\text{eff}}}{\partial g_{ab}} = \frac{\partial V_{\text{eff}}}{\partial \sigma}\bigg|_{\sigma=\overline{\sigma}} \, \frac{\partial \overline{\sigma}}{\partial g_{ab}}
\end{eqaed}
and so forth. Then, to compute the pullback of the metric, on account to~\cref{eq:fermi_RG_line} one need only contract the $(ab)$ indices with the constant tangent vector $v$. This yields a complicated, but explicit, expression for the pullback
\begin{eqaed}\label{eq:pullback_metric}
    G_{vv} \equiv G_{\mathbf{g}(u)}(v,v)
\end{eqaed}
along the RG trajectory in terms of the quantities
\begin{eqaed}\label{eq:pullback_needs}
    \overline{\sigma} \, , \quad \text{grad} (\overline{\sigma})(v) \, , \quad \text{Hess}(\overline{\sigma})(v,v)
\end{eqaed}
evaluated along the flow as functions of the parameter $u$. Our task is then to evaluate each of these quantities along the flow for $\epsilon \ll 1$, starting from~\cref{eq:fermi_gap} and its derivatives. More precisely, only for the $\epsilon$-expansion the limit we work with is $\frac{1}{\epsilon} \gg M \, , k \, \gg N \gg 1$. The hierarchy $M \, , k \, \gg N$ is the one leading to the infinite-distance limit, and in this regime the bulk interpretation of the theory lies on shakier grounds. However, this is not ultimately an issue, for a number of reasons: the relevant case for the holographic picture that we have in mind is $d=3$, for which the information distances are computed numerically without assuming $M \gg N$. We use the $\epsilon$-expansion as an analytical tool to obtain closed-form expression that we can compare qualitatively to the relevant case. Finally, as we shall see, the numerical computations fit the pattern of the $\epsilon$-expansion even outside of the hierarchy $M \gg N$. We will elaborate on this point in~\cref{sec:distances} and~\cref{sec:HS_holography}.

The crucial non-trivial step is finding an asymptotic expression of the gap $\overline{\sigma}(\mathbf{g}(u))$ along the flow. Once obtained, it can be substituted in the first and second derivatives of~\cref{eq:fermi_gap} to solve for the gradient and Hessian in~\cref{eq:pullback_needs}. The latter step is manageable, albeit cumbersome. To approach the former, we begin by observing that only two components of the gap contribute to the metric in~\cref{eq:pullback_metric}, since only a bicritical $2 \times 2$ submatrix of couplings varies along each step of the flow. Writing out the two gap equations from~\cref{eq:fermi_gap}, one finds
\begin{eqaed}\label{eq:eps-exp_fermi_gap}
    \sigma_1 & \sim \frac{\epsilon}{2} \left( \sigma_1 \, \log\left(1 + \frac{1}{\sigma_1^2}\right) + \frac{u}{k} \, \sigma_2 \, \log\left(1 + \frac{1}{\sigma_2^2}\right) \right) , \\
    \sigma_2 & \sim \frac{\epsilon}{2} \left( (1-u) \, \sigma_2 \, \log\left(1 + \frac{1}{\sigma_2^2}\right) + u \, \sigma_1 \, \log\left(1 + \frac{1}{\sigma_1^2}\right) \right) , \\
\end{eqaed}
so that the solutions $\overline{\sigma}_1 \, , \, \overline{\sigma}_2$ are
\begin{eqaed}\label{eq:_eps-exp_fermi_gap_sol}
    \overline{\sigma}_1 \sim \overline{\sigma}_2 \sim e^{-1/\epsilon} \, .
\end{eqaed}
This result mirrors the well-known exponentially suppressed gap in the ordinary Gross-Neveu model.

Armed with~\cref{eq:_eps-exp_fermi_gap_sol}, one can proceed differentiating~\cref{eq:fermi_gap} to solve for the gradient and Hessian perturbatively in $\epsilon$, thus obtaining the pullback $G_{vv}$ of the metric in~\cref{eq:pullback_metric}. Let us stress that this procedure applies equally well to the bosonic case, where however we where able to obtain the fully perturbative metric. In the fermionic case, the non-perturbative nature of the gap complicates matters considerably, and a straightforward expression of $G$ for general couplings is not available. The final result is then independent of $M$ and reads
\begin{equation}
\begin{split}
    & \pi\,e^{2/\epsilon} \epsilon \, G_{vv} \overset{\epsilon \ll 1}{\sim} \\
    &\frac{2 (k+1)N}{((k+1)N-u-1)^2}-\frac{((k+1)N)^2}{((k+1)N-2) ((k+1)N-1) ((k+1)N-u-1)} \\
    &-\frac{((k+1)N-2) ((k+1)N-1)}{((k+1)N-u-1)^3} -\frac{((k-1)N)^2}{(n-1) (1-u) (kN-1) ((k+1)N-2)} \\
    &-\frac{kN^3 \left(-(k^2+1) N + u k N ((k+1)N-2)+k+1\right)}{(N-1) (kN-1) ((k+1)N-1) (N(k N u (u-2)+k+1)-1)} \,.
\end{split}
\end{equation}
Until this point, this is an exact result in $N,k$, expanded only in $\epsilon$.
Further expanding this result in $N,k$ with the hierarchy $k \gg N \gg 1$, we find that the first three terms are subleading with respect to the last two, and keeping only the leading orders one arrives at
\begin{eqaed}\label{eq:large-N_fermi_eps_Gvv}
    \pi\,e^{2/\epsilon} \epsilon \, G_{vv} \overset{\frac{1}{\epsilon} \gg k \gg N \gg 1}{\sim} \frac{1-N u}{1-N (2-u) u}-\frac{1}{N (1-u)} \sim \frac{1-N (1-u) u}{(1-u) (1 - N (2-u) u)} \, .
\end{eqaed}
In the next section we discuss how to use this result, together with its bosonic counterpart, to compute information distances and study infinite-distance limits.

\section{Computing distances}\label{sec:distances}

To summarize, in the preceding sections we have presented the computations of large-$N$ information metrics for the four cases under scrutiny, namely the multicritical bosonic (resp. fermionic) model in $d=3$ and $d=4-\epsilon$ (resp. $d=2+\epsilon$) for $\epsilon \ll 1$.

In the bosonic case we have access to the full metric, numerically in $d=3$ and perturbatively in the $\epsilon$-expansion. The same numerical technique works also for the fermionic case in $d=3$, but the $\epsilon$-expansion about $d=2$ is considerably more difficult compared to the bosonic case. This is due to the fact that, while in the bosonic case the leading-order solution to gap equation is perturbative, up to logarithmic terms, the leading-order term in the fermionic gap is non-perturbative, of the form $\sigma \sim e^{-\frac{1}{\epsilon}}$. In order to circumvent this, instead of computing the full metric (in a neighbourhood of the Gaussian point $\mathbf{g}=0$) $G_{\mathbf{g}}$, we computed its pullback $G_{vv}(u) \equiv G_{\mathbf{g}(u)}(v,v)$ on the RG trajectory that we consider. The length of the RG trajectory represents the \emph{single-step} distance $\Delta^{\text{bdry}}_\text{ss}$ on the boundary side of the holographic correspondence, and is given by
\begin{eqaed}\label{eq:single-step_cft_distance}
    \Delta^{\text{bdry}}_\text{ss} = \int_0^1 \sqrt{G_{vv}(u)} \, du \, .
\end{eqaed}
Generalizing the holographic correspondence between metrics from the case of a conformal manifold to the information metric, one is thus naturally led to propose the bulk distance~\cite{Baume:2020dqd}
\begin{eqaed}\label{eq:single-step_bulk_distance}
    \Delta^{\text{bulk}}_\text{ss} \sim \frac{1}{\sqrt{C_T}} \int_0^1 \sqrt{G_{vv}(u)} \, du
\end{eqaed}
at large $C_T$ central charge. The rationale behind the asymptotics in~\cref{eq:single-step_bulk_distance} is that dividing by $C_T$ is strictly speaking only unambiguous when $C_T$ does not vary along the curve, \emph{i.e.}\ on a conformal manifold. Hence, in the spirit of keeping our proposal as close as possible to this well-established case, the RG trajectories that provide single steps ought to be the ones along which $C_T$ varies the least.

In order to probe large distances in theory space, as we have anticipated one ultimately need consider the hierarchy $M \gg N \gg 1$, and whenever $\frac{1}{\epsilon} \gg 1$ it is also the largest parameter in the hierarchy. We do this within the $\epsilon$-expansion in order to obtain more manageable closed-form expressions for the large-distance behavior, but our numerical computations do not rely on this assumption. In the large-$N$ limit, $C_T$ does not vary to leading order over a single step, and~\cref{eq:single-step_bulk_distance} is well-defined. This is consistent with the intuition that only the large-$N$ limit of the theories we discuss seem to describe sensible gravitational physics in the bulk. However, since we are interested in the leading $\frac{1}{N}$ corrections (to anomalous dimensions of higher-spin currents), picking the trajectories where the \emph{subleading} part of $C_T$ varies the least seems to be the most sensible option. The general structure is~\cite{Petkou:1995vu, Diab:2016spb}
\begin{eqaed}\label{eq:C_T_subleading}
    C^{G}_T \sim a \, N + b
\end{eqaed}
for a single, decoupled critical sector with symmetry group $G = O(N)$ or $U(N)$, where $a \, , \, b = \mathcal{O}(1)$ are known constants~\cite{Petkou:1995vu, Diab:2016spb}. For a free sector, $b=0$ instead. Therefore, summing over the decoupled group factors
\begin{eqaed}\label{eq:C_T_combined}
    C^{\prod_i G_i}_T = \sum_{i} C_T^{G_i} \sim \sum_{i} \left( a \, N_i + b_i \right) ,
\end{eqaed}
where $b_i = b$ for critical factors and $b_i = 0$ for free factors.

In order to simplify the computation of the information distance, we consider simple multicritical models of the form $O(N)^M$ or $U(N)^M$, \emph{i.e.}\ with fixed equal ranks $N \gg 1$. The higher-spin limit then amounts to taking $M \to \infty$, since, as we will discuss in detail in~\cref{sec:HS_holography}, the anomalous dimensions $\gamma_{\text{HS}}$ of the higher-spin currents of interest vanishes according to
\begin{eqaed}\label{eq:anomalous_dim_HS_ratio}
 \frac{\gamma^{O(NM)}_{\text{HS}}}{\gamma^{O(N)^M}_{\text{HS}}} \, , \, \frac{\gamma^{U(NM)}_{\text{HS}}}{\gamma^{U(N)^M}_{\text{HS}}} \sim \frac{\mathcal{O}(1)}{M} \, . 
\end{eqaed}
We are interested in expressing~\cref{eq:anomalous_dim_HS_ratio} in terms of the information distance along the chain of single-step flows connecting the initial and final fixed points. According to the preceding discussions, the chain of RG flows in~\cref{eq:bosonic_bicritical_flow} and~\cref{eq:fermionic_bicritical_flow} comprises $M-1$ steps, and the full bulk distance is then the sum over single steps of~\cref{eq:single-step_bulk_distance}, namely
\begin{eqaed}\label{eq:full_bulk_distance}
    \Delta^{\text{bulk}} \sim \sum_{k=1}^{M-1} \Delta^{\text{bulk}}_k
\end{eqaed}
where the $k$-th step connects
\begin{eqaed}\label{eq:kth_step}
    & O((k+1)N) \times O(N)^{M-k-1} \quad \longrightarrow \quad O(kN) \times O(N)^{M-k} \qquad \text{bosonic} \, ,\\
    & U(kN) \times U(N)^{M-k} \quad \longrightarrow \quad U((k+1)N) \times U(N)^{M-k-1} \qquad \text{fermionic} \, .
\end{eqaed}
In the following we derive expressions for $\Delta^\text{bulk}(M)$ in order to check whether it diverges as $M \to \infty$ and, if so, how~\cref{eq:anomalous_dim_HS_ratio} depends on it in this limit.

\subsection{Bosonic case}\label{sec:bosonic_dist_results}

\subsubsection{Distance in \texorpdfstring{$d=4-\epsilon$}{d=4-epsilon}}\label{sec:bosonic_dist_eps-exp}

\begin{figure}[ht!]
    \centering
    \includegraphics[width=\textwidth]{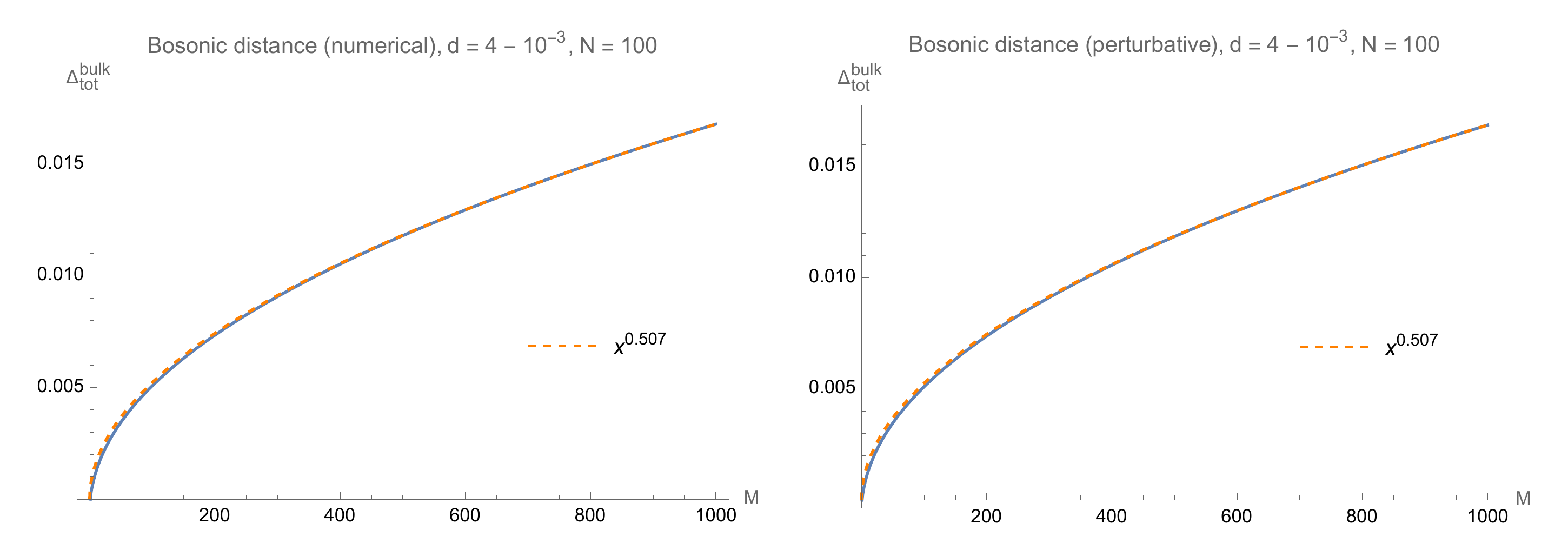}
    \caption{Bosonic distance for $d = 4 - 10^{-3}$, $N = 100$ and $M \leq 1000$. The numerical computation (left panel) agrees with the perturbative prediction (right panel). The square-root behaviour is fitted with a power-law model.}
    \label{fig:bose_4d}
\end{figure}

\begin{figure}[ht!]
    \centering
    \includegraphics[scale=0.7]{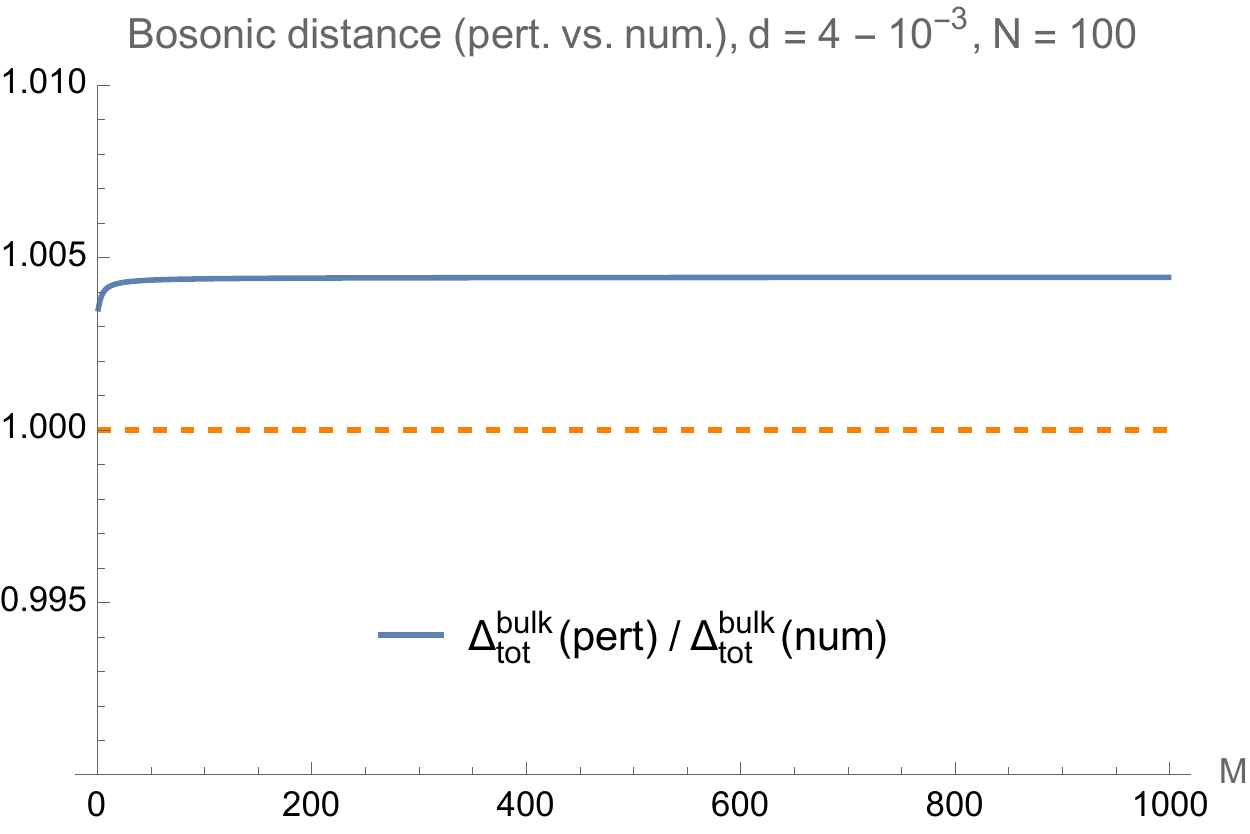}
    \caption{Ratio of the bosonic distances for $d = 4 - 10^{-3}$ and $M \leq 1000$, computed using the full numerical model and the small $\epsilon$, large-$M$ perturbative analysis. The difference between the two approaches gives an error less than $5\times 10^{-3}$.}
    \label{fig:bose_4d_ratio}
\end{figure}

In the $\epsilon$-expansion for the bosonic model the asymptotics of the metric can be computed explicitly from~\cref{eq:pert_free_energy_r=0}, but its expression is quite long and cumbersome. However, for $k \gg N \, \gg 1$ the pullback of~\cref{eq:pullback_metric} greatly simplifies to
\begin{eqaed}\label{eq:bose_eps-exp_Gvv}
    G_{vv} \sim \frac{\epsilon^2}{2 \pi^2 \, N
    } \, \log\left(\frac{4}{\epsilon^2 (1-u)u}\right) .
\end{eqaed}
The single-step bulk distance is therefore independent of $k$ and scales as $1/\sqrt{M}$ due to the factor of $C_T$ in~\cref{eq:single-step_bulk_distance}. The total bulk distance thus scales like $\sqrt{M}$, or more precisely
\begin{eqaed}\label{eq:bose_eps-exp_tot_dist}
    \Delta^{\text{bulk}} \sim \frac{\epsilon}{N} \, \sqrt{\log \frac{1}{\epsilon}} \, \sqrt{M}
\end{eqaed}
up to an irrelevant numerical factor. Thus, the distance diverges as $M \gg 1$, which is a proper infinite-distance limit. We have compared the prediction $\Delta^{\text{bulk}} \sim \mathcal{O}(1) \, \sqrt{M}$ with a numerical computation, and the resulting very good agreement is depicted in~\cref{fig:bose_4d}. Importantly, the agreement does not rely on $M \gg N$. Examining the ratios, one finds a relative discrepancy of less than $5 \times 10^{-3}$ for $M \leq 10^3$, as shown in~\cref{fig:bose_4d_ratio}.

\subsubsection{Distance in \texorpdfstring{$d=3$}{d=3}}\label{sec:bosonic_dist_3d}

\begin{figure}[ht!]
    \centering
    \includegraphics[scale=0.7]{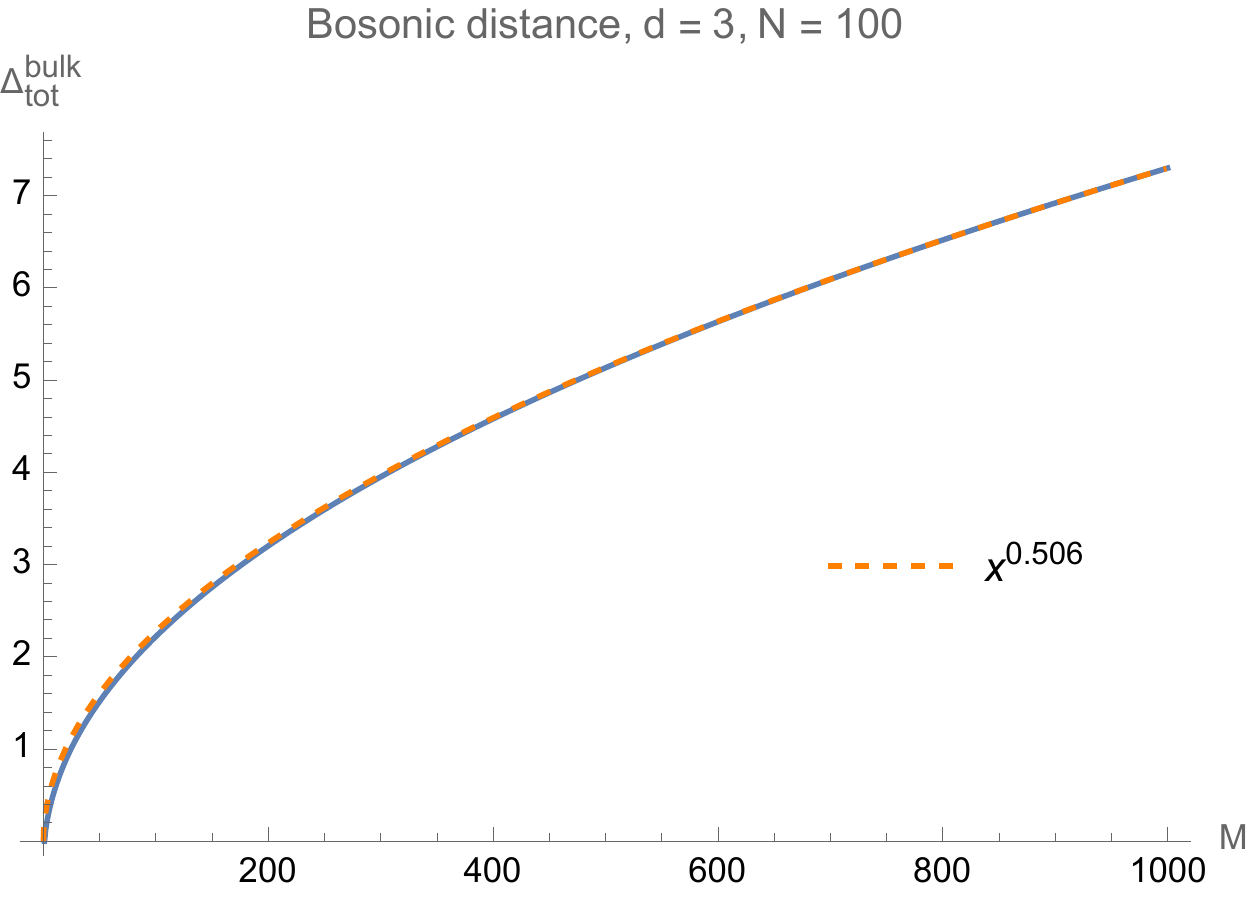}
    \caption{Bosonic distance for $d=3$, $N = 100$ and $M \leq 1000$. The square-root behaviour is fitted with a power-law model.}
    \label{fig:bose_3d}
\end{figure}

For $d=3$, arguably more interesting for higher-spin holography, the perturbative approch breaks down and we relied on a numerical method. In more detail, we numerically solved~\cref{eq:gap_eqs} for couplings along the trajectory, and subsequently computed the pullback metric of~\cref{eq:pullback_metric}. Finally, we evaluated the total distance summing over single steps according to~\cref{eq:full_bulk_distance}. For $M \leq 1000$ and $N=100$ the results are shown in~\cref{fig:bose_3d}, where a polynomial fit shows a behavior seemingly compatible with the square-root result from the $\epsilon$-expansion, once again without assuming $M \gg N$, so that the regime of validity of a bulk interpretation for $d=3$ has an overlap with the power-like scaling of the distance. Indeed, the fit in~\cref{fig:bose_4d} also shows similar variance for the exponent in the fit.

\subsection{Fermionic case}\label{sec:fermionic_dist_results}

\subsubsection{Distance in \texorpdfstring{$d=2+\epsilon$}{d=2+epsilon}}\label{sec:fermionic_dist_eps-exp}

In the fermionic model, the $\epsilon$-expansion exhibits exponentially suppressed gaps. This means that numerical methods are considerably less reliable with respect to the bosonic case. In particular for, say, $\epsilon=10^{-3}$ the numerical analysis completely breaks down due to underflow. Fortunately, for $\epsilon \ll 1$ we can derive a closed-form expression for the distance starting from~\cref{eq:large-N_fermi_eps_Gvv}.

We cannot straightforwardly expand~\cref{eq:large-N_fermi_eps_Gvv} in $N$, since combinations such as $N u$ can be arbitrarily small within the integration domain $u \in [0, 1]$. In particular, zeros and poles of $G_{vv}$ as a function of $u$ depend on $N$. For finite $N$, this is a function of $u$ that has a positive and a negative part, with zeros located at $\frac{1}{2} \left(1 \pm \sqrt{\frac{N-4}{N}}\right)$ and poles at $1-\sqrt{1 - \frac{1}{N}}$ and $1$. $G_{vv}$ is positive between the two zeros. Furthermore, when $N \to \infty$, all zeros and poles are pushed to $0$ or $1$ and we have a smooth positive definite function of $u \in [0,1]$, which we can express by expanding the expression of $G_{vv}$ in powers of $N$ for $u \in \left[\frac{1}{2} - \frac{1}{2} \sqrt{\frac{N-4}{N}}, \frac{1}{2} + \frac{1}{2} \sqrt{\frac{N-4}{N}}\right]$,
\begin{equation}\label{eq:fermi_eps_final_Gvv}
    G_{vv} \sim \frac{e^{-2/\epsilon}}{\pi \, \epsilon} \frac{1}{2-u} \,.
\end{equation}
When $N \to \infty$, this is the leading contribution to $G_{vv}$ defined on $(0,1)$, and indeed the integral of~\cref{eq:large-N_fermi_eps_Gvv} (defined with some determination of the square root) has a vanishing imaginary part for $N \to \infty$.
Thus, the single-step integrated bulk distance is
\begin{eqaed}\label{eq:fermi_2d_ss_dist}
    \Delta^\text{bulk}_k \sim \frac{1}{\sqrt{C_T}} \int_0^1 \sqrt{G_{vv}} \, du \sim \frac{e^{-1/\epsilon}}{\sqrt{\epsilon}} \frac{1}{\sqrt{N M}}
\end{eqaed}
up to an irrelevant numerical factor. For the full trajectory $U(N)^M \to \cdots \to U(MN)$, the integrated bulk distance is therefore
\begin{eqaed}\label{eq:fermi_2d_tot_dist}
    \Delta^\text{bulk} = \sum_{k=1}^{M-1} \Delta^\text{bulk}_k \sim \frac{e^{-1/\epsilon}}{\sqrt{\epsilon}} \sqrt{\frac{M}{N}} \, ,
\end{eqaed}
again up to an irrelevant numerical factor. Notice that both closed-form expressions for $\epsilon \ll 1$ feature the same $\sqrt{M}$ scaling for $M \gg 1$, which is therefore an infinite-distance limit also in the fermionic case. Let us observe that the $\epsilon$-dependence of~\cref{eq:fermi_2d_tot_dist} and~\cref{eq:bose_eps-exp_tot_dist} are related by the substitution $\epsilon_{\text{Bose}} = e^{-1/\epsilon_{\text{Fermi}}}$. This can be traced back to the $\epsilon$-dependence of the solutions $\overline{\sigma}$ to the respective gap equations in~\cref{eq:bose_eps-exp_gap_sol} and~\cref{eq:eps-exp_fermi_gap}, which are also related in this fashion.

\subsubsection{Distance in \texorpdfstring{$d=3$}{d=3}}\label{sec:fermionic_dist_3d}

\begin{figure}[ht!]
    \centering
    \includegraphics[scale=0.7]{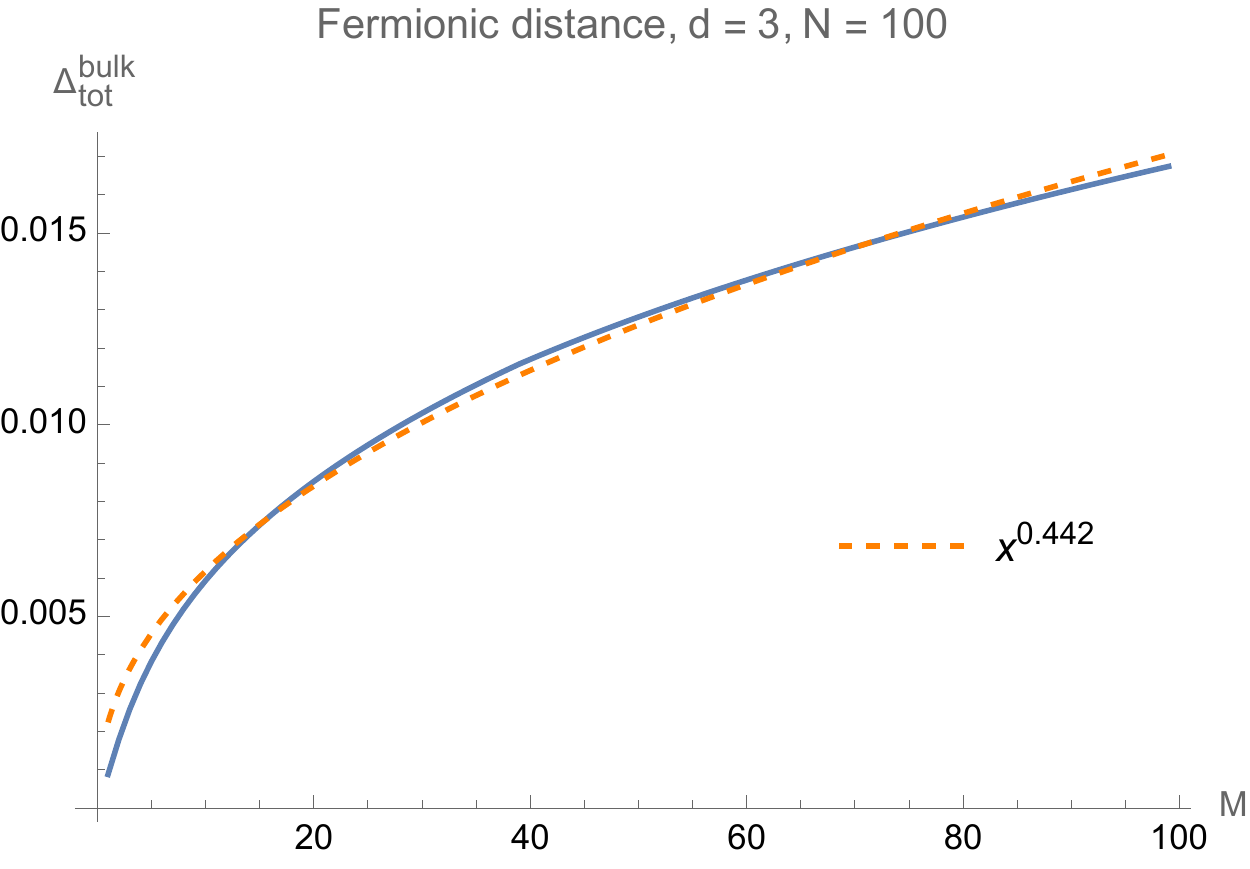}
    \caption{Fermionic distance for $d = 3$, $N = 100$ and $M \leq 100$. The square-root behaviour is fitted with a power-law model. The accuracy of the fit is limited by the range $M \leq 100$, over which our method shows numerical instabilities.}
    \label{fig:fermi_3d}
\end{figure}

In $d=3$ once again numerical methods are the only tool at our disposal. However, the fermionic case is still quite numerically unstable even at $\epsilon=1$, presumably because of residual issues stemming from the exponential underflows at $\epsilon \ll 1$. Our numerical analysis shows that the computations are stable and reliable up to $M \approx 120$, which is not particularly large. However, the plot in~\cref{fig:fermi_3d} still shows a clear polynomial trend, which seems compatible with a large-$M$ square-root behavior considering the larger uncertainty of this fermionic computation. Moreover, as in the bosonic case, the power-like scaling appears to hold also in the regime of validity of a bulk interpretation.

\subsection{Chern-Simons-matter CFTs}\label{sec:cs_matter}

In order to build a sensible bulk interpretation of (multi)critical vector models, the standard procedure involves introducing a singlet constraint on spectra. In $d=3$ this can be implemented by coupling vector matter to a Chern-Simons sector~\cite{Aharony:2011jz, Giombi:2011kc, Aharony:2012ns} at positive level $k_{\text{CS}} \gg 1$. In the large-$N$ limit with
\begin{eqaed}\label{eq:lambda_CS}
   \lambda \equiv \frac{N}{k} \equiv \frac{N}{k_\text{CS} + N} \in (0,1)
\end{eqaed}
fixed, one finds a line of CFTs for all $\lambda$, and higher-spin symmetry is present at the edges of the interval, where the theory is equivalent to the critical/free bosons or free/critical fermions. Inside the interval, higher-spin currents are not conserved as operators, but their divergence may vanish inside certain correlators~\cite{Maldacena:2012sf,Giombi:2016zwa}, and therefore it is natural to ask whether the information distance in $\lambda$-space behaves according to the expectations from the distance conjecture. For a $U(N)$ gauge group, the leading-order free energy on the sphere $S^3$ reads~\cite{Klebanov:2011gs}
\begin{eqaed}\label{eq:CS_free_energy}
   F_\text{CS} = \frac{N}{2} \, \log(k+N) - \, \sum_{j=1}^{N-1} (N-j) \, \log \left( 2 \, \sin \frac{\pi j}{k+N}\right) ,
\end{eqaed}
The role of these theories in the context of the CFT distance conjecture has been outlined in~\cite{Perlmutter:2020buo}, but since $\lambda$ is not a continuous parameter for any finite $N$ one cannot rely on the Zamolodchikov metric. However, the information metric can still be defined, since flows connecting different values will result in the same distance. This is due to the fact that there is only one possible simple curve connecting two points on a one-dimensional manifold. From this perspective it thus seems sensible to promote $\lambda$ to a continuous parameter.\footnote{An alternative approach to promote a discrete parameter to a continuous one is discussed in~\cite{Stout:2022phm}.}

Since we are interested in the intensive metric of the coupled Chern-Simons-matter theory, the infinite-radius limit of the metric arising from~\cref{eq:CS_free_energy} is well-defined. Out of the two potentially singular regions, only $\lambda \ll 1$ is accessible from~\cref{eq:CS_free_energy}, since
\begin{eqaed}\label{eq:CS_asymp_F}
    F_\text{CS} \overset{\frac{1}{\lambda} , N \gg 1}{\sim} \frac{N^2}{2} \, \log \frac{1}{\lambda}
\end{eqaed}
is the dominant contribution to the total free energy. The behavior in the complementary regime $\lambda \to 1^-$ would then be determined by duality. Since the action is linear in $1/\lambda$, the second derivative of $-F_\text{CS}$ with respect to $1/\lambda$ yields the asymptotic metric
\begin{eqaed}\label{eq:CS_metric}
   ds^2 & \overset{\frac{1}{\lambda} , N \gg 1}{\sim} \lambda^2 \, d\left(\frac{1}{\lambda}\right) \otimes d\left(\frac{1}{\lambda}\right) = \frac{d\lambda \otimes d\lambda}{\lambda^2}
\end{eqaed}
up to a prefactor. Thus, the distance to $\lambda = 0$ is infinite. By duality, one can argue that the same conclusion applies to $\lambda = 1$. This result also fits with our general expectations due to how free theories and higher-spin symmetry are recovered at the endpoints.

More in detail, the distance $\Delta(\lambda , \lambda_0) \sim \, c \, \log\frac{\lambda_0}{\lambda}$ diverges logarithmically in $\lambda$, where $\lambda < \lambda_0$ and $c > 0$ is some positive constant. The anomalous dimensions of higher-spin currents scale according to~\cite{Giombi:2016zwa}
\begin{eqaed}\label{eq:CS_HS_anomalous_dims}
   \gamma_s \overset{N \gg 1}{\sim} \frac{\pi \lambda}{2N \sin(\pi \lambda)} \left(a_s \, \sin^2\left(\frac{\pi \lambda}{2}\right) + \frac{b_s}{4} \, \sin^2(\pi \lambda) \right) ,
\end{eqaed}
where $a_s$ and $b_s$ only depend on the spin. For $\lambda \ll 1$ one finds
\begin{eqaed}\label{eq:CS_HS_anomalous_dims_small_lambda}
   \gamma_s \overset{\frac{1}{\lambda} , N \gg 1}{\sim} \frac{(a_s+b_s)\pi^2}{8N} \, \lambda^2 \, ,
\end{eqaed}
so that, in terms of the distance $\Delta$,
\begin{eqaed}\label{eq:CS_SDC}
   \frac{\gamma_s(\lambda)}{\gamma_s(\lambda_0)} \sim e^{- \frac{2}{c} \, \Delta(\lambda, \lambda_0)} \, . 
\end{eqaed}
Hence, in the case of the Chern-Simons theory space parametrized by $\lambda$, the recovery of higher-spin symmetry seems to be exponentially fast in the distance. Consistently with our preceding considerations, one may be led to speculate that the matrix-like nature of Chern-Simons degrees of freedom, which dominates over the vector-like one in this case, be responsible for the stringy exponential decay of higher-spin anomalous dimensions. The presence of a gauge theory can also be detected via the large spin asymptotic of the higher-spin anomalous dimensions, which is $\log s$, see~\cite{Alday:2007mf}. 

\section{Higher-spin holography}\label{sec:HS_holography}

Let us briefly summarize the main features of the higher-spin/vector model holography. We will concentrate on the spectrum, \emph{i.e.}\ on the dictionary between single-trace operators and bulk fields, most of which is a simple consequence of representation theory. The most basic example was anticipated~\cite{Flato:1978qz} long before the AdS/CFT correspondence and concerns free vector models. Two important twists of the story are to add critical vector models to the scene~\cite{Klebanov:2002ja,Sezgin:2003pt,Leigh:2003gk,Bekaert:2012ux} and to extend all of the models to Chern-Simons vector models~\cite{Giombi:2011kc,Giombi:2012ms, Giombi:2016ejx}. 

In the simplest case we have a $O(N)$ or $U(N)$ vector $\phi^i$ of free scalars. The free scalar field corresponds to a unitary irreducible representation of the conformal symmetry algebra $so(d,2)$, say $V$. Then, it is a well-known result~\cite{Flato:1978qz, Craigie:1983fb} that all quasi-primary single-trace operators (we do not display the tensor indices of the various operators and only indicate their number)
\begin{align}
    J_{s}&= \bar{\phi}^i \partial^s \phi_i + \dots \,, && \Delta=d+s-2\,,\\
    J_0&= \bar{\phi}^i\phi_i\,, && \Delta=d-2\,,
\end{align}
are covered by a scalar operator $J_0$ and by conserved tensors $J_s$ with $s=1,2,3,\dots$ for the $U(N)$ case and with $s=2,4,6,\dots$ for the $O(N)$ case. From the representation theory point of view this corresponds to the decomposition of $V\otimes V$ or $(V\otimes V)_S$ into irreducible representations of the conformal algebra. The standard AdS/CFT dictionary
\begin{align}
    \partial^m J_{mabc\,\dots}&=0 &&\Longleftrightarrow && \delta \Phi_{\mu_1\,\dots\,\mu_s}=\nabla_{(\mu_1}\xi_{\mu_2\,\dots\,\mu_s)}
\end{align}
implies that the dual theory has gauge (for $s>0$) fields $\Phi_{\mu_1\,\dots\,\mu_s}$ with the same range of $s$, \emph{i.e.}\ all spins for the dual of the free $U(N)$ model and all even spins for the dual of the $O(N)$ one. 

Similarly, one can take a $U(N)$ vector multiplet $\psi_i$ of free fermions. In $d=3$ the spectrum of the single-trace operators is very close to the one of the free scalar CFT:
\begin{align}
    J_{s}&= \bar{\psi}^i \gamma\partial^{s-1} \psi_i + \dots\,, && \Delta=s+1\,,\\
    J_0&= \bar{\psi}^i\psi_i\,, && \Delta=2\,.
\end{align}
The latter property is a first sign of the three-dimensional bosonization duality~\cite{Giombi:2011kc, Maldacena:2012sf, Aharony:2012nh,Aharony:2015mjs,Karch:2016sxi,Seiberg:2016gmd}. Another comment is that all CFTs with higher spin conserved tensors in $d\geq3$ are free CFTs~\cite{Maldacena:2011jn,Boulanger:2013zza,Alba:2013yda,Alba:2015upa}, possibly, in disguise.

A simple extension of this construction is to take several vector multiplets, $\phi^i_a$, that transform under $U(N)\times U(M)$ or $O(N)\times O(M)$ and to impose the singlet constraint with respect to $U(N)$ or $O(N)$ only. The higher spin currents then get decorated with $U(M)$ or $O(M)$ indices
\begin{align}
    U(N)&: & J_{s}{}^a{}_b&= \bar{\phi}^a_i \partial^s \phi_b^i + \dots \,, \\
    O(N)&: & J_{s}{}^{ab}&= {\phi}^a_i \partial^s \phi^b_j \delta^{ij} + \dots \,.
\end{align}
In the case of $U(N)\times U(M)$ the single-trace operators take values in the adjoint of $U(M)$ and for the case of $O(N)\times O(M)$ the currents with odd/even spins take values in the adjoint/rank-two symmetric plus singlet representations of $O(M)$. Since $U(M)$ or $O(M)$ is the leftover global symmetry, it becomes gauged in the bulk and one expects fields $\Phi_{\mu_1\,\dots\,\mu_s}$ to take values in the same representations of $U(M)$ or $O(M)$. In particular, the spin-one field always carries the adjoint representation. Likewise, one can decorate currents $J_{s}{}^a{}_b$ built out of $U(N)
\times U(M)$ free fermions $\psi_i^a$ with $U(M)$. A thorough discussion of all  free cases from the representation theory point of view as well as supersymmetric extensions can be found in~\cite{Konstein:1989ij}.  

Further generalizations of the basic dualities discussed so far can be obtained with the help of critical vector models~\cite{Klebanov:2002ja,Leigh:2003gk}. A quartic interaction $(J_0)^2$ can be added and tuned to criticality. In the large-$N$ limit the only visible effect for the single-trace operators is in that the dimension $\Delta$ of $J_0$ jumps from $1$ to $2+\mathcal{O}(1/N)$ for the bosonic vector models and from $2$ to $1+\mathcal{O}(1/N)$ for the fermionic ones (also called Gross-Neveu model). As is well-known~\cite{Klebanov:1999tb} the bulk effect of such double-trace deformations is in the change of the boundary conditions on the dual field $\Phi_0$. 

While the microscopical origin of the singlet constraint is still mysterious in $d>3$, at least in $d=3$ there is a natural mechanism:\footnote{It was discussed already in~\cite{Klebanov:2002ja} that as long as $4-\epsilon$ expansion allows one to access the Wilson-Fisher fixed point on the CFT side it may be some sense in which higher-spin gravities can exist in $\rm{AdS}_{5-\epsilon}$. At least one-loop corrections to the free energy of higher-spin gravities can be computed in $\rm{AdS}_{5-\epsilon}$~\cite{Skvortsov:2017ldz}, but we do not pursue this idea further in the present paper.} one can gauge the $U(N)$ or $O(N)$ symmetry by also adding a Chern-Simons term at level $k$ to the action. In the $k\rightarrow \infty$ limit the Chern-Simons terms effectively decouples while imposing the singlet constraint (see, however, the discussion of the free energy puzzle in \cite{Giombi:2013fka,Giombi:2016pvg,Gunaydin:2016amv}). Moreover, the Chern-Simons coupling brings in one additional parameter, the level $k$. In the large-$N$ limit the effective coupling is $\lambda=N/k$, which is almost a continuous parameter. The conjecture of the three-dimensional bosonization duality is that free/critical bosonic/fermionic vector models coupled to Chern-Simons are pairwise identical as CFT's~\cite{Giombi:2011kc, Maldacena:2012sf, Aharony:2012nh,Aharony:2015mjs,Karch:2016sxi,Seiberg:2016gmd}, \emph{e.g.}\ the critical bosonic vector model with the Chern-Simons interaction is dual to the free fermion model coupled to the Chern-Simons term. It should be interesting to explore similar dualities in the multicritical models that we have studied in this paper, as well as their symplectic counterparts which could have an interpretation in terms of dS/CFT~\cite{Anninos:2011ui}.

For a given CFT to have a well-defined, semi-classical, local holographic description with finitely many fields certain conditions are to be met (in some region of the parameter space). Vector models violate several of the usual assumptions: there is no large gap in the spectrum and, hence, the dual theory has to have infinitely many fields from the start; the structure of correlation functions in free and critical large-$N$ vector models imply that the dual higher-spin gravity has to be very non-local~\cite{Bekaert:2015tva,Maldacena:2015iua,Sleight:2017pcz,Ponomarev:2017qab} and, as a result, cannot be constructed with the help of the usual field theory techniques.\footnote{It is worth mentioning that there exists a consistent and local truncation of the higher-spin gravity that is dual to vector models~\cite{Ponomarev:2016lrm,Skvortsov:2018uru,Sharapov:2022awp}, Chiral higher-spin gravity~\cite{Metsaev:1991mt,Metsaev:1991nb,Ponomarev:2016lrm}. It should be dual to a closed subsector of Chern-Simons vector models. } A reconstruction, \emph{i.e.}\ fitting the most general Ansatz for bulk interactions with the already known correlation functions, should be possible~\cite{Bekaert:2015tva} and can, at least formally, be performed directly under the path integral for vector models, see \emph{e.g.}~\cite{deMelloKoch:2018ivk,Aharony:2020omh} and references therein.\footnote{As a side remark one can also mention proposals for other higher-spin gravities with propagating degrees of freedom~\cite{Segal:2002gd,Tseytlin:2002gz,Bekaert:2010ky,Hahnel:2016ihf,Adamo:2016ple},~\cite{Sperling:2017dts,Tran:2021ukl,Steinacker:2022jjv},~\cite{Metsaev:1991mt,Metsaev:1991nb,Ponomarev:2016lrm,Skvortsov:2018uru,Sharapov:2022awp} and topological ones~\cite{Blencowe:1988gj,Bergshoeff:1989ns,Campoleoni:2010zq,Henneaux:2010xg,Pope:1989vj,Fradkin:1989xt,Grigoriev:2019xmp,Grigoriev:2020lzu} that do not immediately fit the vector model holography we discuss in the present paper. Meanwhile, all these theories are perturbatively local in contrast to vector models' duals.}

One subtle point regarding the interplay between the weakly gauged $O(N)$ symmetry on the CFT side and the leftover global $O(M)$ symmetry that becomes gauged in the AdS-dual gravitational description is that $M$ should not exceed $N$, for the bulk fields must be independent. Indeed, let us consider the anti-symmetrized product of $n$ bulk fields
\begin{align}
    & (\phi^{i_1}_{[a_1} \phi^{j_1}_{|b_1|}\delta_{i_1 j_1})\,\dots\,(\phi^{i_n}_{a_{n}]} \phi^{j_n}_{b_{n}}\delta_{i_n j_n}) && \Longleftrightarrow && \Phi_{[a_1 |b_1|}\,\dots\,\Phi_{a_{n}]b_{n}} \,,
\end{align}
for which we also wrote down the dual operator. Clearly, the expression will vanish for $n>N$, but it has no reasons to vanish on the bulk side for $M \geq n$. Recalling that the bulk coupling $G$ is of order $1/N$ we have some bounds on the size of the Yang-Mills group $M$, $G^{-1}\gtrsim M$. In other words, for $M$ large enough the large-$N$ expansion is inapplicable. Similar subtleties were also pointed out \emph{e.g.}\ in~\cite{Das:2003vw,Aharony:2020omh} and studied recently in~\cite{Aharony:2022feg}.

\section{Implications for the swampland}\label{sec:swampland_implications}

\begin{figure}[ht!]
    \centering
    \includegraphics[scale=0.7]{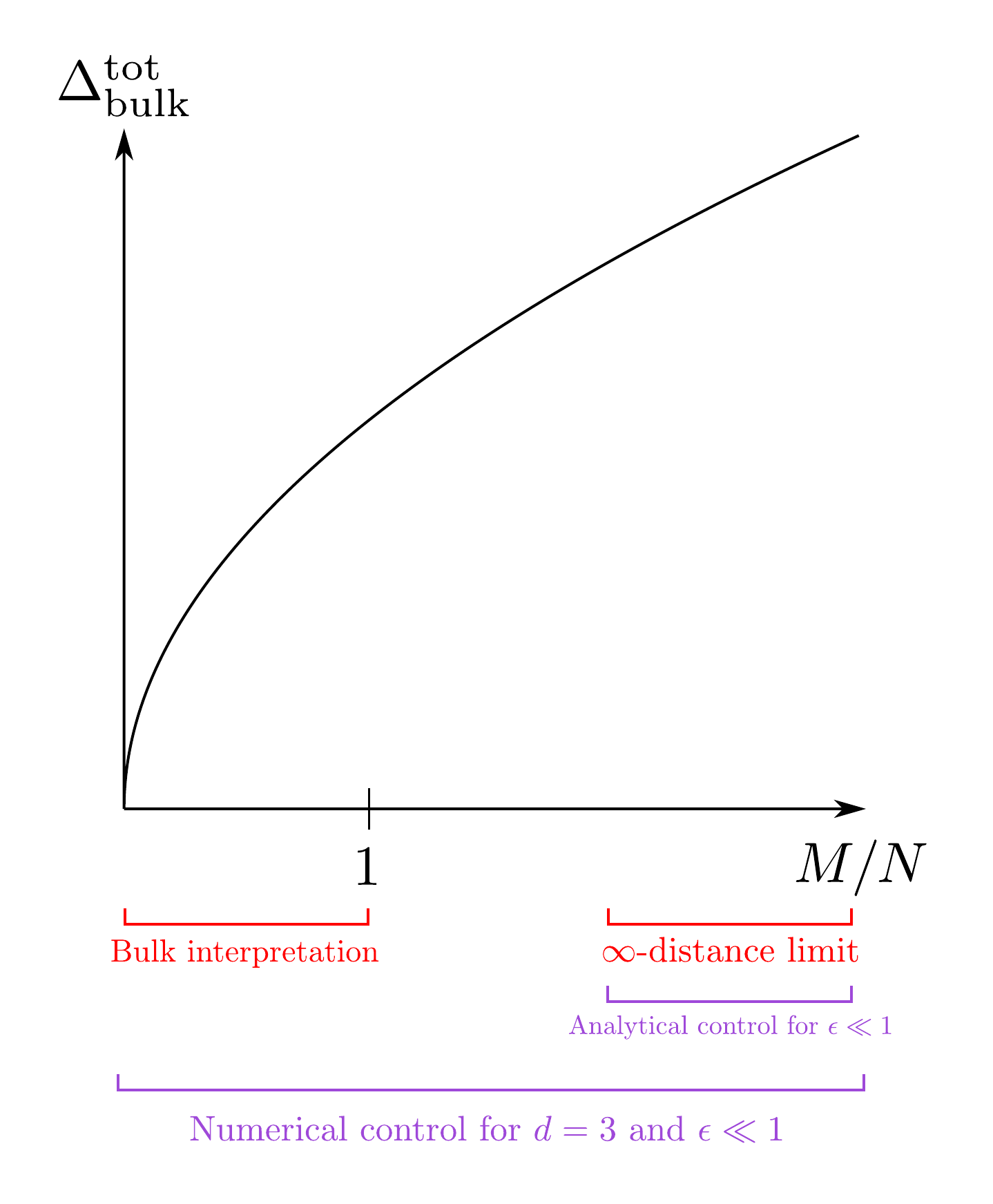}
    \caption{The power-like behavior $\Delta \sim \sqrt{M}$, or at least $\Delta \sim M^\alpha$ with $\alpha \approx 0.5$, persists down to small steps $M \lesssim N$, where a bulk description is more reliable due to the absence of the extra constraints discussed in~\cref{sec:HS_holography} (see also~\cite{Aharony:2022feg}). The analytic control given by the $\epsilon$-expansion provides a convenient cross-check, but our ultimate interest is in the case $d=3$, and our numerical technique does not require $M \gg N$.}
    \label{fig:delta_plot}
\end{figure}

The results that we have collected suggest a natural interpretation in terms of the swampland distance conjecture~\cite{Ooguri:2006in}, in particular regarding its CFT counterpart~\cite{Baume:2020dqd, Perlmutter:2020buo}. The RG trajectories that we have considered always include a $O(N)$ or $U(N)$ symmetry subgroup (we will refer to the former for brevity). It is thus tempting to impose a singlet constraint on this subgroup to identify duals of higher-spin states in the bulk: at the decoupled $O(N)^M$ fixed point the anomalous dimensions of the corresponding higher-spin currents scale as $\gamma_\text{HS} \sim \frac{1}{N}$. Gauging a diagonal $O(N)$ subgroup, for sufficiently large Chern-Simons level the above discussion is unaffected, with the proviso that only one stress tensor is now conserved, allowing a more sensible holographic interpretation.\footnote{In the presence of many independent conserved spin-2 tensors, the union of corresponding AdS spaces glued at their conformal boundary has been proposed as a holographic description~\cite{Aharony:2006hz, Apolo:2012gg}.} To elaborate further on this point, in the preceding section we have discussed how a proper bulk interpretation seems to require $M \lesssim N$, while the infinite-distance limit requires $M \gg N$. This is not unexpected, since infinite-distance limits encode a breakdown of the bulk EFT description, while its boundary counterpart can remain perfectly sensible. We nonetheless find the same power-like behavior of the distance $\Delta(M)$ as a function of the number of steps $M$, as depicted in~\cref{fig:delta_plot}.

The chain of RG flows gradually enhances the fixed points to $O(MN)$, along a distance $\Delta \sim \sqrt{M}$. The anomalous dimensions gradually \emph{decrease} up to $\gamma_\text{HS} \sim \frac{1}{MN}$. Thus, in the infinite-distance limit one has
\begin{eqaed}\label{eq:anomalous_HS_dims_result}
   \frac{\gamma^{O(NM)}_{\text{HS}}}{\gamma^{O(N)^M}_{\text{HS}}} \, , \, \frac{\gamma^{U(NM)}_{\text{HS}}}{\gamma^{U(N)^M}_{\text{HS}}} \sim \frac{\mathcal{O}(1)}{\Delta^2} \, .
\end{eqaed}
This power-like recovery of higher-spin symmetry is strikingly different from the exponential decay that is apparently ubiquitous in string-theoretic constructions, and to our knowledge~\cref{eq:anomalous_HS_dims_result}  is the first example of this kind. Of course, care must be taken in placing this result in the context of the swampland distance conjecture, since distances outside of ordinary moduli spaces have been explored to a much lesser extent~\cite{Stout:2021ubb, Basile:2022zee, Stout:2022phm}. Furthermore, let us remark that alternative definitions of the distance at stake are possible, namely applying the information distance(s) in different ways compared to what we have done in this paper~\cite{Stout:2022phm}.\footnote{We thank John Stout for pointing this out.} Nevertheless, our approach based on RG flows connecting discrete sets of (fixed) points may be motivated by additional physical considerations. Namely, much like moving in moduli spaces corresponds to changing VEVs with no energy cost, moving in a discrete landscape of vacua corresponds to crossing domain walls, whose existence is entailed by the cobordism conjecture in the presence of gravity~\cite{McNamara:2019rup}. Holographically, these processes are described by RG flows interpolating between CFTs. This construction does not promote parameters from discrete to continuous ones, and indeed in stringy constructions the former typically arise from Dirac-quantized fluxes, number of fundamental (brane) sources and so on.

Finally, it is unclear to what extent \emph{exponential} decay is crucial in order to place a theory in the swampland. In particular, it would be interesting to perform a similar computation in the setting of ABJ triality~\cite{Chang:2012kt}, where a string-theoretic picture is available. This specific example, as well as general considerations in holography, highlight the potential origin of the power-like vs exponential discrepancy, namely the difference between vector-like and matrix-like holographic theories. We began analyzing the issue in the context of Chern-Simons-matter CFTs, but it seems reasonable that the same principle accounts for different scalings in large-$N$ limits as well. To wit, computing the same information distance along RG flows in $\mathcal{N}=4$ Super Yang-Mills and its dual supergravity leads to an exponential decay of Kaluza-Klein masses in the large-$N$ limit~\cite{Basile:xxx}. It is conceivable that a scenario in which one can interpolate between the two behaviors would feature higher-spin anomalous dimensions decaying with a steeper and steeper power of the distance, eventually turning to an exponential decay in the matrix-like limit. At any rate, defining and studying infinite-distance limits in discrete landscapes of stabilized vacua is of considerable interest in phenomenological model building.

In the most extreme scenario, in which only strictly exponential decays are consistent and the construction that we have proposed is a correct extension of well-established swampland principles, the models that we have discussed in this paper would be placed in the swampland, or at least excluded from the landscape of string-theoretic constructions. In the opposite extreme, in which our construction has nothing to say about the swampland, one can regard our analysis as a test of an extension of Conjecture III as put forth by~\cite{Perlmutter:2020buo} in purely CFT terms. Alternatively, one can regard this as a study of how higher-spin gravity behaves differently from typical stringy expectations. 

All in all, the main takeaway that we would like to emphasize is that the infinite towers of light states that characterize higher-spin gravity indeed appear at suitably defined large distances in the theory space of their holographic duals. On the other hand, the specific exponential behavior of masses and anomalous dimensions is deeply tied to stringy or matrix-like degrees of freedom. Considerations of quantum-gravitational consistency of non-stringy models based on distance conjectures should thus take this into account. As a final comment, we stress that there may even be settings in which a light tower of states appears at a finite distance. We provide a concrete example of this sort in~\cref{sec:gff} where, on the other hand, the price to pay is the absence of a local energy-momentum tensor, in line with the considerations of~
\cite{McNamara:2020uza}.

\section*{Acknowledgements}
A.C.\ and E.S.\ are Research Associates of the Fund for Scientific Research - FNRS, Belgium. S.P.\ is a FRIA grantee of the Fund for Scientific Research - FNRS, Belgium. The work of A.C, S.P. and I.B.\ was supported by FNRS under Grants No.\ F.4503.20 (``HighSpinSymm'') and T.0022.19 (``Fundamental issues in extended gravitational theories''). The work of E.S. was partially supported by the European Research Council (ERC) under the European Union’s Horizon 2020 research and innovation programme (grant agreement No 101002551) and by FNRS under Grant No.\ F.4544.21.

\noindent We would like to thank Thomas Basile, Florent Baume, José Calderón-Infante, John Stout and Niccolò Risso for helpful discussions. 

\appendix

\section{Generalized free fields and compact theory spaces}\label{sec:gff}
In the preceding sections we have explored theories where higher-spin points lie at infinite information distance. This nicely fits with the expectations coming from the distance conjecture, even in the generalized setting of this paper. As we have remarked, one of the most striking features of the analyses of~\cref{sec:bose_models} and~\cref{sec:fermi_models} is the power-like scaling of the higher-spin anomalous dimensions with the information distance that we used. It is thus natural to ask whether there exist cases in which higher-spin points do not lie at infinite distance.

In this appendix we examine generalized free fields as a simple example of calculable theory space with a non-trivial metric, and we find that the theory space is compact with respect to the information distance that we used in this paper.

Concretely, we consider the momentum-space Euclidean action
\begin{eqaed}\label{eq:gff_action}
    S = \frac{C_\Delta}{2} \int \frac{d^dp}{(2\pi)^d} \, \phi(-p) \, \abs{p}^{d - 2\Delta} \, \phi(p) \, .
\end{eqaed}
The inverse propagator $G_\Delta(p) = C_\Delta \, \abs{p}^{2\Delta - d}$ contains the scaling dimension $\Delta = \frac{d-2}{2} + \gamma$ as a free parameter. Here we define generalized free fields in terms of~\cref{eq:gff_action}, rather than in terms of OPEs, since it is more convenient in order to compute the information metric. As we shall see shortly, this treatment highlights how non-locality enters the discussion on infinite-distance limits.

The action in~\cref{eq:gff_action} depends on $\Delta$ in a non-linear fashion. Hence, according to~\cref{eq:information_metric_def}, the information metric reads
\begin{eqaed}\label{eq:gff_metric}
    g_{\Delta \Delta} = \langle \partial_\Delta \partial_\Delta S \rangle + \partial_\Delta \partial_\Delta \log Z
\end{eqaed}
up to a volume factor, where a UV regulator $\Lambda$ is implicit. Multiple decoupled generalized free scalars would simply yield a diagonal metric $g_{\Delta_i \Delta_j} = g_{\Delta_i \Delta_i} \delta_{ij}$. From~\cref{eq:gff_metric}, a quick computation gives
\begin{eqaed}\label{eq:gff_metric_computed}
    g_{\Delta \Delta} = \int^\Lambda \frac{d^dp}{(2\pi)^d} \left( \frac{C_\Delta'^2}{2C_\Delta^2} - \frac{2C_\Delta'}{C_\Delta} \log \frac{\abs{p}}{\Lambda} + 2 \log^2 \frac{\abs{p}}{\Lambda} \right) \, ,
\end{eqaed}
where $C'_\Delta \equiv \partial_\Delta C_\Delta$. 

The higher-spin limit $\Delta \to \frac{d-2}{2}$ depends on the behavior of $C_\Delta$. However, it seems that for any reasonable normalization $C_\Delta$ the metric would be regular near the higher-spin point, since $C_{\Delta = \frac{d-2}{2}} \equiv 1$ for canonically normalized free fields. The absence of metric singularities would then imply that the distance $\int \sqrt{g_{\Delta \Delta}} \, d\Delta$ to the higher-spin point be finite.

If one were to trust the information metric of~\cref{eq:information_metric_def} to extend the distance conjecture beyond the standard settings, this result may appear thorny at first glance. However, it is actually compatible with the considerations of~\cite{Perlmutter:2020buo} on locality. In order to have a sensible theory of gravity in the bulk side of a holographic correspondence, the locality of the stress tensor seems to be an important requirement. Thus, it seems reasonable to conclude that this example does not constitute a proper counterexample to the (web of) distance conjecture(s). Let us conclude briefly mentioning that the locality of a holographic dual in quantum gravity also plays a crucial role in other aspects of the swampland program, as highlighted in~\cite{McNamara:2020uza}.

\printbibliography

\end{document}